\documentclass{aa} 

\usepackage{amsmath,amsfonts,amssymb}
\usepackage[dvipsnames]{xcolor}
\usepackage{color}
\usepackage{url}
\usepackage[varg]{txfonts}
\usepackage{graphicx}
\usepackage{enumitem} 
\usepackage{array}
\usepackage{dblfloatfix}
\usepackage{multicol}
\usepackage[normalem]{ulem} 

\usepackage{natbib,twoopt}
\usepackage[breaklinks=true]{hyperref} 
\bibpunct{(}{)}{;}{a}{}{,} 
\makeatletter
\newcommandtwoopt{\citeads}[3][][]{\href{http://adsabs.harvard.edu/abs/#3}%
{\def\hyper@linkstart##1##2{}%
\let\hyper@linkend\@empty\citealp[#1][#2]{#3}}}
\newcommandtwoopt{\citepads}[3][][]{\href{http://adsabs.harvard.edu/abs/#3}%
{\def\hyper@linkstart##1##2{}%
\let\hyper@linkend\@empty\citep[#1][#2]{#3}}}
\newcommandtwoopt{\citetads}[3][][]{\href{http://adsabs.harvard.edu/abs/#3}%
{\def\hyper@linkstart##1##2{}%
\let\hyper@linkend\@empty\citet[#1][#2]{#3}}}
\newcommandtwoopt{\citeyearads}[3][][]%
{\href{http://adsabs.harvard.edu/abs/#3}
{\def\hyper@linkstart##1##2{}%
\let\hyper@linkend\@empty\citeyear[#1][#2]{#3}}}
\makeatother

\usepackage{color}
\hypersetup{colorlinks=true,linkcolor=blue,citecolor=blue,urlcolor=blue}

\usepackage[breaklinks=true]{hyperref} 

\makeatletter
\renewcommand*\aa@pageof{, page \thepage{} of \pageref*{LastPage}}
\makeatother

\newcommand{\sop}[1]{{\color{black} #1}}

\usepackage{xcolor}

\begin{document} 

\title{3D magneto-hydrodynamical simulations of stellar convective noise for improved exoplanet detection}

\subtitle{I. Case of regularly sampled radial velocity observations}

 \author{S. Sulis \inst{1,2,3}
 \and D. Mary \inst{3}
 \and L. Bigot \inst{3} 
 }
 \institute{    
 Aix Marseille Univ, CNRS, CNES, LAM, Marseille, France\label{inst1}\\
\email{sophia.sulis@lam.fr}
 \and
 Space Research Institute, Austrian Academy of Sciences, Schmiedlstr. 6, 8042 Graz, Austria\label{inst2}
  \and
 Universit\'e C\^ote d'Azur, Observatoire de la C\^ote d'Azur, 
                CNRS, Lagrange UMR 7293, CS 34229, 
 06304, Nice Cedex 4, France\label{inst3}
 }

 \date{Accepted}

 \abstract
 {Convective motions at the stellar surface generate a stochastic colored noise source in the radial velocity (RV) data. This noise impedes the detection of small exoplanets. Moreover, the unknown statistics (amplitude, distribution) related to this noise make it difficult to estimate the false alarm probability (FAP) for exoplanet detection tests.
 }
 {In this paper, we investigate the possibility of using 3D magneto-hydrodynamical simulations (MHD) of stellar convection to design detection methods that can provide both a reliable estimate of the FAP and a high detection \sop{power}.
 } 
 {We tested the \sop{realism} of 3D simulations in producing solar RV \sop{by comparing them} with the observed disk integrated velocities taken by the GOLF instrument on board the SOHO spacecraft. We \sop{presented} a new detection method based on periodograms standardized by these simulated time series, applying several detection tests to these standarized periodograms.
 }
 {The power spectral density of the 3D synthetic convective noise is consistent with solar RV observations for short periods.
 For regularly sampled observations, the analytic expressions of FAP derived for several statistical tests applied to the periodogram standardized by 3D simulation noise are accurate.  
The adaptive tests considered in this work (Higher-Criticism, Berk-Jones), which are new in the exoplanet field, may offer better detection performance than classical tests (based on the highest periodogram value) in the case of multi-planetary systems and planets with eccentric orbits. 
 } 
 {3D MHD simulations are now mature enough to produce reliable synthetic time series of the convective noise affecting RV data. These series can be used to access to the statistics of this noise and derive accurate FAP of tests that are a critical element in the detection of exoplanets down to the cm.s$^{-1}$ \sop{level}. 
 }


 \maketitle

\section{Introduction}

 At the time of this writing, $880$ extrasolar planets have been discovered so far by the radial velocity (RV) technique\footnote{Source: \url{exoplanet.eu}, confirmed planets (01/2020).}. 
In this sample, $52~\%$ of these consist of planets that are more massive than Jupiter and $14~\%$ that have a mass inferior to $10$ Earth-masses ($M_\oplus$). Since the detection of HD215152c \citepads{2011arXiv1109.2497M}, only $19$ planets have been found with a mass $\le 2 M_\oplus$ and all of the latter are short-period ($\le 50$ days) planets orbiting stars that are less massive than the Sun. 

Indeed, detecting planets is easier around low-mass stars (as the ratio of the planet to stellar masses is higher) and, thus, a first strategy consists in monitoring cool M dwarfs to increase the detection probability. This has been the purpose of recent surveys with spectrographs such as CARMENES \citepads{2014SPIE.9147E..1FQ} and SPIRou \citepads{2017haex.bookE.107D}.
On the other hand, new instruments such as ESPRESSO \citepads{2010SPIE.7735E..0FP} and EXPRES \citepads{2016SPIE.9908E..6TJ} have been developed to ensure the long-term stability that is needed to detect signals of terrestrial planets orbiting main sequence G-dwarf stars (with an amplitude around $10$-$30$ cm.s$^{-1}$).

However, detecting planet signatures at the cm.s$^{-1}$ level remains  challenging as spurious Doppler shifts of various origins may dominate the RV series.
The activity at the surface of the host star is one of the main sources generating changes in depth, width, and asymmetries of the absorption lines.
Disentangling the planetary signal from the stellar activity ``noise'' is an active research topic (see e.g., \citeads{2007A&A...473..983D, Aigrain_2012, Haywood_2014, Lagrange_2010,Meunier_2017b, 2018AJ....156..180W,2018A&A...620A..47D, Chaplin_2019}; and references therein) 
and stellar activity has already led to several controversial planet detections in the past (e.g., $\alpha$CenB b,  \citeads{2012Natur.491..207D,Hatzes_2013,2016MNRAS.456L...6R}, GJ581 d and g, \citeads{2010ApJ...723..954V,2014Sci...345..440R}, GJ667 c, and f \citeads{Anglada2013,2014ApJ...793L..24R}). 

This activity results from the contribution of various phenomena, which can be classified as a function of their  correlation timescales. For main-sequence Solar-like stars, \sop{the three main noise sources originate from}: \sop{1)} \textit{cyclic} stellar oscillation eigenmodes (a few minutes), \sop{2)} \textit{stochastic} surface convection motions (min-hrs), \sop{and 3)} \textit{(quasi) periodic} stellar activity -- spots, plages, flares -- modulated with the stellar rotation or cycle (days-years). We note that even if \sop{active regions} are more frequent at the maximum phases of the cycle, they can have lifetimes that are shorter than the rotation period \citepads{1997ApJ...485..319S}.

In this paper, we aim to consider how the influence of the stochastic noise due to stellar convective motion \sop{could be counteracted}. We ignore other sources of stellar RV variations (e.g., oscillations and active regions) and consider them as already corrected in our time series (e.g., through activity-sensitive lines \citepads{1995ApJ...438..269B,2018AJ....156..180W} or dedicated filtering technique \citepads{Chaplin_2019}).

Convective noise can significantly alter the detection of exoplanets at the sub-m.s$^{-1}$ level \citepads{Meunier_2015,Meunier_2017a, Meunier_2019,Cegla_2019}.
The main technique proposed so far for mitigating its contribution in the RV series down to some tenth of cm.s$^{-1}$ consists of averaging several (typically \sop{two or three}) measurements of a target star during a night and separating them by at least \sop{two} hours \citepads{2011A&A...525A.140D, 2019MNRAS.487.1082C}. However, the convection acts as a correlated noise over timescales longer than \sop{two-to-four} hours (and even longer for supergranulation) and some correlations remain by using this observational procedure \citepads{Meunier_2015}. Moreover, this technique is performed at the cost of a small number of data points per night, leading to a critical lack of knowledge of the statistical properties for stellar activity as a whole.
Other methods for dealing with convection noise consist of modeling the stellar activity as a correlated noise when fitting for RV planetary Keplerian signatures. Examples of common empirical models that we can find in the literature are red (i.e., power law) noise \citepads{2014MNRAS.437.3540F}, moving averaged noise \citepads{2014MNRAS.441.1545T}, or Gaussian processes \citepads{10.1093/mnras/stv1428}.
In practice, these empirical modelings should be used with caution as their results may lead to different conclusions, as shown in a recent RV challenge \citepads{2017A&A...598A.133D}.

In this study, we question the reliability of traditional methods for determining the statistical significance of the detection in the presence of correlated noise. This significance is based on the value of the false alarm probability (FAP) of statistical tests (see review in \citeads{khan2017}). Traditionally, the FAP is derived under the assumption that the noise within the data (or the data residuals) is an uncorrelated white Gaussian noise (WGN). 
In this work, we propose a new method to access the significance of the detection of (quasi-)periodic signals in the presence of a correlated noise, providing that we can generate reliable (non-parametric) time series of this noise. 
We propose to use state-of-the-art 3D magneto-hydrodynamical (MHD) simulations of stellar surfaces to generate the noise series.
We note that such simulations have already been investigated to determine the impact of convection on exoplanet detection \citepads{Cegla_2013,Cegla_2018,Cegla_2019b}.
Our analysis focuses on evaluating the reliability of MHD simulations in reproducing the time series of solar convective noise and on investigating the statistical benefit of using such simulated RV for deriving accurate FAP. In this paper, the benefit of using these simulations in the detection process is described for the case of regularly sampled time series. The case of  an irregular sampling will be developed in a second paper\footnote{The reader can refer to \citetads{2017arXiv170606657S} for preliminary indications about how this work can be extended to the case of irregularly sampled observations.}, whereas the present analytical studies can nevertheless provide a useful proxy of the performance that can be expected in the 
case of an irregular sampling close to regular (e.g., one point per night at roughly the same hour).  
 
This paper is divided into five sections. 
In Sec.~\ref{Sec2}, we evaluate the realism of the 3D MHD simulations. In Sec.~\ref{Sec3}, we use the  standardized periodogram and present several detection tests to exploit this periodogram. In particular, we discuss some tests that are new to the exoplanet field: namely, the Higher-Criticism \citepads{donoho2004} and the Berk-Jones tests \citepads{Berk1979}. In Sec.~\ref{Sec4}, we perform a numerical study to investigate the benefit of our procedure and present our conclusions in Sec.~\ref{Sec5} and \ref{Sec6}.

\section{Simulated solar granulation noise}
\label{Sec2}

As a preamble, we aim to test the \sop{realism} of 3D MHD simulated RV time series of convective (granulation) noise and compare it to the RV time series obtained using the spectrophotometer \textit{Global Oscillation at Low Frequencies} (GOLF).

\subsection{Space measurements of RV solar convective noise}

\label{Sec21}

Measurements from spaceborne missions represent an excellent opportunity to validate the simulated velocities of solar convection. Indeed, they are not affected by the alternation of day and night or any ground-based follow-up problems (e.g., the influence of the Earth's atmosphere) and provide regularly sampled time series at high cadence. 

Since 1996, the GOLF spectrophotometer on board the \textit{Solar and Heliospheric Observatory} (SoHO) spacecraft takes an almost continuous measurement of the solar disk-integrated position of the Sodium doublet. More particularly, it measures the position in the ``blue'' and ``red'' wings of the lines at $\pm~108$ \AA~ from the center of the lines, which are located at $\lambda = 5895.924$ \AA~ (${\cal D}_1$) and $5889.950$ \AA~ (${\cal D}_2$). 
The solar light enters into a sodium vapor cell and a magnetic field splits the absorption lines (Zeeman effect). 
Then the Doppler shift (i.e., velocity) is evaluated as the flux ratio on these two points of the lines' wings \citepads[see p. 328,]{Unno_1989}:
\begin{equation} 
        v(t) \propto ~\frac{F_B(t)-F_R(t)}{F_B(t)+F_R(t)},
        \label{eq_vspectra} 
\end{equation}
where $F_B$ and $F_R$ are the fluxes in the blue and red wings, respectively. For more technical details about this velocity extraction, we refer  to \citetads{1993ExA.....4...87B, 1995SoPh..162...61G, Garcia_2005} and \citetads{2018A&A...617A.108A}. 

After roughly one year of GOLF measurements, an instrumental failure happened and the velocity extraction was done using only one side of the sodium doublet: the blue wings (where the solar intensity comes from the bottom of the photosphere) from 1996 to 1998 and from 2002 until now and the red wings (where the solar intensity comes from the upper layers of the photosphere) between these dates \citepads{Garcia_2005}. 
Therefore, a careful calibration of the GOLF data was needed to obtain consistent velocities and several calibrations have been proposed. We chose to use the recent level-2 GOLF data\footnote{\url{www.ias.u-psud.fr/golf/templates/access.html}} calibrated as described in \citetads{2018A&A...617A.108A}. 
In order to have the same sampling as in our MHD simulations, we sampled the GOLF time series every minute\footnote{The original sampling was $20$ seconds.}. 
Moreover, we divided the GOLF time series into two-day sequences to study the RV correlations on the granulation timescales (from a few minutes up to several hours) and to validate them with Monte Carlo (MC) simulations based on a large number of solar subseries (\sop{see} the statistical results presented in Sec.~\ref{Sec3}).
From the entire sample of two-day sequences, we removed the ones containing observation gaps to have a perfectly regularly sampled time series (as our working hypothesis throughout this paper). 

Finally, we applied a low-pass filter of $1620$ $\mu$Hz (i.e. ,$10.3$ minutes) passband to filter out the oscillations modes and to restrict sensitivity to pick up only the convective noise. 
We computed the velocity root-mean-square (rms) of each $182$ sequences available on the $1996$ dataset (i.e., at solar cycle minimum, no calibration problem) and obtained an average value of $49$ cm.s$^{-1}$, which is in agreement with \citetads{1999ASPC..173..297P}. 
An example of a two-day sequence and the corresponding periodogram will be shown in Sec.~\ref{Sec2_compa}.

\subsection{Synthetic time series of the RV solar convective noise}
\label{Sec22}

\subsubsection{Magneto-hydrodynamical simulations of the solar surface}
\label{Sec221}

We use the state-of-the-art radiative MHD code ({\scriptsize STAGGER CODE}, 
\citeads{Nordlund_1995}) to simulate the surface convection and stratification of the Sun. In a 3D local-box model of the solar atmosphere (size: ${\rm 8000\times8000\,kms}$ and $+500$ and $-3400$ km above and below the surface at optical depth $\tau=1$), the code solves the full set of conservative MHD equations coupled to an accurate treatment of the radiative transfer. The horizontal sizes of the domain are defined to contain a sufficient number of granules at each time-step. The code is based on a sixth-order explicit finite difference scheme. The equations are solved on a staggered mesh where the thermodynamic variables are cell-centered, while the flux is shifted to the cell edge. The domain of simulation contains the entropy minimum located at the surface (photosphere) and is extended deep enough to have a flat entropy profile at the bottom (adiabatic regime). The code uses periodic boundary conditions horizontally and opened boundaries vertically. At the bottom of the simulation, the inflows have constant entropy and pressure. The outflows are not constrained and are free to pass through the boundary.
We used a realistic equation-of-state that accounts for ionization, recombination, dissociation \citepads{1988ApJ...331..815M}, and continuous line opacity \citepads{2008A&A...486..951G}.
Radiative transfer is crucial since it drives convection through entropy losses at the surface \citepads{1998ApJ...499..914S} and is solved using the Feautrier's scheme along with several inclined rays (one vertical, eight inclined) through each grid point. The wavelength dependence of the radiative transfer is taken into account using a binning scheme, in which the monochromatic lines are collected into $12$ bins. The numerical resolution used for the present simulation is $120^3$. The choice of this modest resolution is a compromise between sufficient fine grid to catch enough of the inhomogeneities and sufficiently small to minimize the computing and storing costs of very long-run simulation. 
 The solar parameters that define our 3D model are $\rm T_{eff} = 5775\pm 30$ K, $\log g = 4.44$ and a solar chemical composition \citepads{Asplund_2009}. The uncertainty in $\rm T_{eff}$ represents the fluctuations due to convection and p-modes. The average magnetic field in our simulation is $\sim 100$ G, as observed by (Hanle) spectropolarimetry \citepads{2004Natur.430..326T}. 

In this work, we use an exceptionally long series of solar snapshots computed to study the properties of solar p-modes \citepads{Bigot_inprep}. It represents $53.14$ days with a sampling of $60$ seconds (i.e., $76~528$ snapshots). To our knowledge, this is the longest series ever generated with such a 3D code. For the present study, we filtered out these modes since they have unrealistic large amplitudes (due to their small inertia) in such shallow boxes of granulation simulation. The synthetic sodium doublet lines are obtained for each snapshot by a monochromatic line transfer within [$5884.000$,$~5901.945$] $~\AA$ and at a resolution of $20~000$. 

The synthetic line intensities $I (t,x,y,\lambda,\mu,\phi)$ and the continuum $C (t,x,y,\lambda,\mu,\phi)$ are computed for each $x$ and $y$, the horizontal Cartesian coordinates of the simulation box and for several inclined rays defined by $\mu$, the cosine of the six limb angles, and four azimuthal angles $\phi$. The chosen discrete $\mu_i$ values are defined by the Gauss-Radau procedure. For six angles, we then have  $\mu_i = \{0.12, 0.39, 0.60, 0.80, 0.92, 1.00\}$.
We averaged these intensities both horizontally and in azimuth to obtain our time-dependent center-to-limb intensity $I(t,\lambda,\mu)$ and continuum $C(t,\lambda,\mu)$, from which we \sop{will} extract the radial velocities \sop{the following sub-sections}.

\subsubsection{RV dependence on the center-to-limb position}
\label{Sec223b}

\begin{figure}[t]
\resizebox{\hsize}{!}{\includegraphics{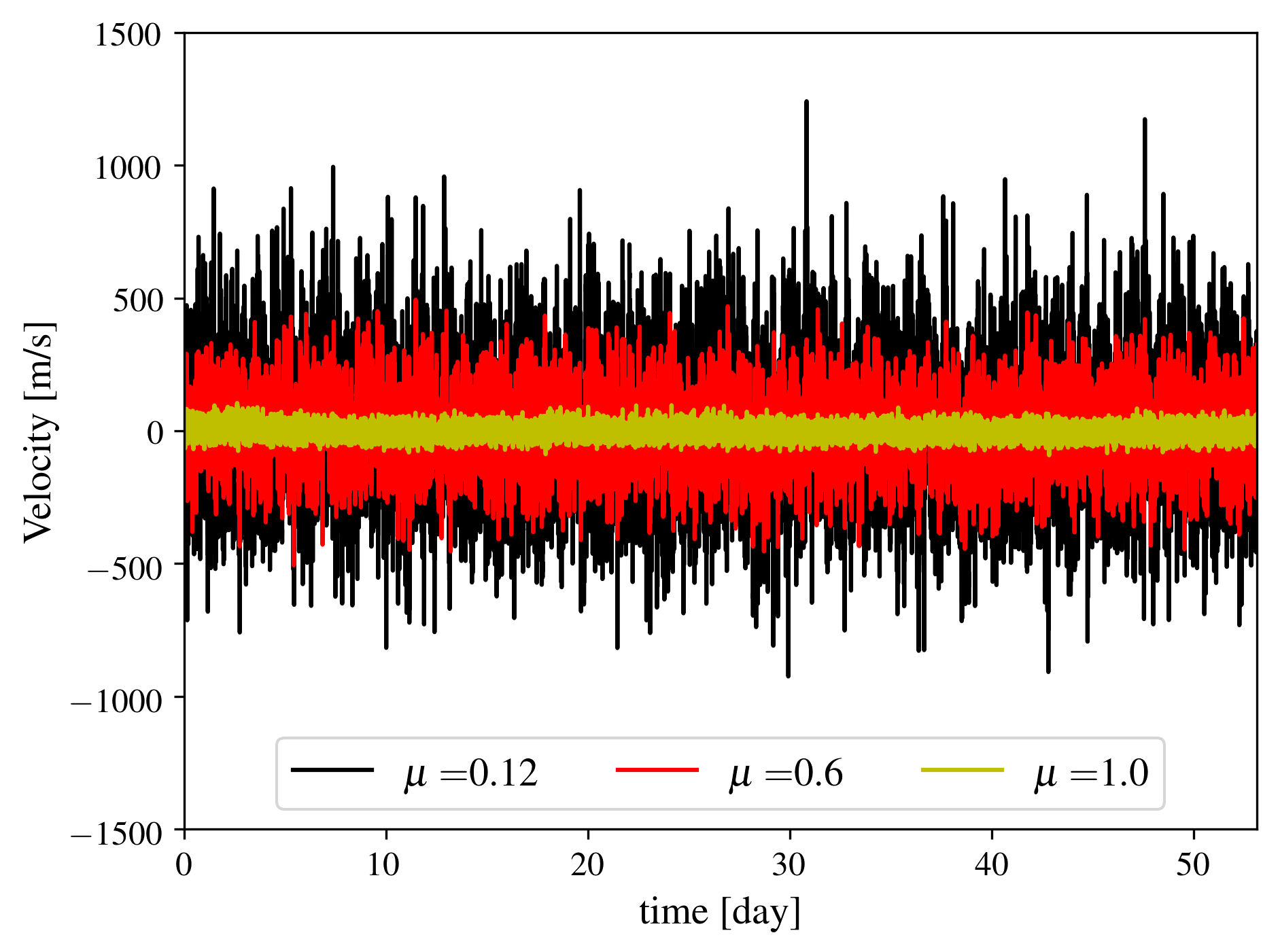}}
\caption{Comparison of velocities time series extracted from the Sodium doublet lines at different $\mu$. For each time series, the oscillation modes have been filtered out. }
\label{Fig_Vmu} 
\end{figure}

The radial velocities associated to each value of $\mu_i$ are obtained using \eqref{eq_vspectra}.
To compute \eqref{eq_vspectra}, we generated the fluxes $F(t,\lambda,\mu_i)$ as the ratio of $I(t,\lambda,\mu_i)$ over $C(t,\lambda,\mu_i)$ for each $\mu_i$. 
We then extracted the mean line profile $F_0(\lambda,\mu_i)$ by averaging the fluxes $F(t,\lambda,\mu_i)$ over $t$ and used this reference profile to evaluate the fluxes ratio involved in \eqref{eq_vspectra}. 
Finally, we translated these Doppler shifts into velocities using a proportional factor ($\kappa$) that results from a Taylor development around the considered wavelength $\lambda_0$ \citepads[see p. 328,]{Unno_1989}. This factor needs to be evaluated for each line of the Sodium doublet. \sop{It writes}:
\begin{equation} 
\frac{1}{\kappa} = \frac{1}{c} ~ \frac{\partial {\rm ln} F(t,\lambda, \mu) }{\partial {\rm ln} \lambda }\Bigg|_{\lambda = \lambda_0},
 \label{eq_kappa} 
 \end{equation}
with $c$ the speed of light and $\lambda_0$ the wavelength corresponding to an intensity level of reference. We set this level of reference to $F_0(\lambda_0^B,\mu_i)=F_0(\lambda_0^R,\mu_i)=0.5$ with $\lambda_0^B$ and $\lambda_0^R$ the wavelengths in the blue and the red wings, respectively. 
The RV time series associated with three of the discrete $\{\mu_i\}$-values are shown in Fig.~\ref{Fig_Vmu}.

The extracted radial velocities are strongly decreasing from the limb to the disk center, with typical rms velocities of $234.2$ m.s$^{-1}$ at the limb ($\mu=0.12$) and $23.3$ m.s$^{-1}$ at the disk center ($\mu=1$), as it is observed for the Sun \citepads{2018A&A...611A...4L}. This is explained by the fact that the observer does not see the same components of convective flows at the limb and the disk center. Indeed, the radial velocity is the projection of the total convective velocity, which includes both the vertical and horizontal velocities. At the disk center, radial and vertical velocities are the same, but at the limb, we only see the horizontal component. Since convection is strongly decelerating and horizontally diverging at the surface so that the gas overturns back to the interior in vertical downflows, the horizontal speeds are much larger than the vertical ones \citepads{1998ApJ...499..914S,2009LRSP....6....2N}. This explains the much larger values found at the limb than at the disk center. Moreover, the contribution of the small vertical velocity at the limb is strictly zero due to the projection effect. Despite the large rms velocities at the limb, we see in the following section that its contribution is limited due to surface projection effect when considering the disk-integrated velocities.

\subsubsection{Disk-integrated RV }
\label{Sec222} 

A single 3D simulation box represents a tiny fraction of the solar surface. Typically, we need $ N_{B}= 2 \pi R_\odot^2/\ell^2 \approx 4.7 \times 10^{4}$ simulation boxes to cover the visible solar disk (i.e., half of the solar surface) with $R_\odot$ the solar radius and $\ell=8$ Mm the horizontal size of the simulation box. 
The difference in RV amplitude between those extracted from one simulation box (with $\mu$-dependent rms velocity $>> 1$ m.s$^{-1}$; see Fig.~\ref{Fig_Vmu}) and the solar disk-integrated observations ($\sim 49$ cm.s$^{-1}$) is due to the cancellation of positive (upflows) and negative (downflows) fluctuations when averaging over the entire disk. The reader might consult \citetads{Ludwig_2006} for an in-depth discussion about this effect in the case of brightness fluctuations (see also \citeads{2008ssma.book.....S}).
To generate a synthetic time series of the solar granulation as seen from disk-integrated observations, we follow a similar methodology to \citetads{Ludwig_2006}, \citetads{Chiavassa_2017} and \citetads{Cegla_2019b} based on our single simulation box. The idea is to use the $76~528$ available synthetic line profiles to patch a surface equivalent to the solar disk. 
Contrary to the previously mentioned studies, which used very short time series of a couple of hours, we have approximately $2\times N_{B}$ boxes to patch the solar disk at a given time, $t$, that allows us to cover the entire disk without duplication of the same snapshots. This allows us to avoid using  the same snapshot in the patching procedure multiple times, which could lead to unavoidable correlations.
In the present study, we randomly distributed the snapshots all over the surface, with the condition that two consecutive patches should correspond to times that are separated by at least $20$ min to minimize possible correlations. 
For each patch $k$, we have an associated emergent intensity $I(t,\lambda,\mu_k)$. The values of $\mu_k$ are calculated depending to the position of the patch on the grid and the intensities $I(t,\lambda,\mu_k)$ interpolated from the six Gauss-Radau values using a second-order polynomial function. Then we let each of these patches evolves independently; that is, for each patch $k$, we performed a new interpolation from the six Gauss-Radau values to derive the new value for $I(t+1,\lambda,\mu_k)$ corresponding to the considered $\mu_k$.
For each $t$ and $\lambda$, we evaluated the disk integrated emergent flux as a function of wavelength,
 \begin{equation} 
 {\cal F}(t,\lambda)= {\sum_{k=1}^{N_B}}\, I(t,\lambda,\mu_k)~\mu_k,
        \label{eq0}
\end{equation}
and we normalized the flux \eqref{eq0} by its corresponding value in the continuum ${\cal F}_C(t,\lambda)= {\sum_{k=1}^{N_B}}\,C(t,\lambda,\mu_k)~\mu_k$.
As in Sec.~\ref{Sec222}, we then generated the mean line profile $F_0(\lambda)$, which is our reference spectral line, to calculate the final Doppler shifts resulting from these disk-integrated synthetic Sodium line spectra. Finally, we extracted the Doppler velocity by measuring the flux ratio in the two points of each of the lines' wings using \eqref{eq_vspectra} with the proportional factor given in \eqref{eq_kappa}.

The acoustic modes are naturally generated by the convective fluctuations inside a simulation box. However, in one shallow box, the modes have much lower inertia than the real observed p-modes. They have therefore much larger amplitudes. Hence, we eliminated their contribution to the RV time series by using a low-frequency filter of $1620$ $\mu$Hz passband (the same applied to GOLF time series; see Sec.~\ref{Sec21}). 
The rms of the final synthetic RV time series is $0.507$ m.s$^{-1}$, which is a value very close to the observed rms from space with GOLF \citepads{1999ASPC..173..297P}, and from the ground with HARPS-N \citepads{2019MNRAS.487.1082C}. We note that our rms value does not take into account the possible contribution of the granulation noise to the high frequencies ($\nu>1620~\mu$Hz) as we filtered them to remove the contribution of the stellar p-modes.

Other rms values due to granulation can be found in the literature. For example, \citetads{Meunier_2015} derived an rms that is twice higher ($80$ cm.s$^{-1}$), \citetads{Cegla_2012} derived a similar value ($40$ cm.s$^{-1}$) and \citetads{Cegla_2019b} derived a smaller value ($10$ cm.s$^{-1}$). The latter authors discuss the influence of the magnetic field that can reduce the velocity of the granulation flows in the 3D simulations. 

 \begin{figure*}[t] 
 \resizebox{\hsize}{!}{\includegraphics{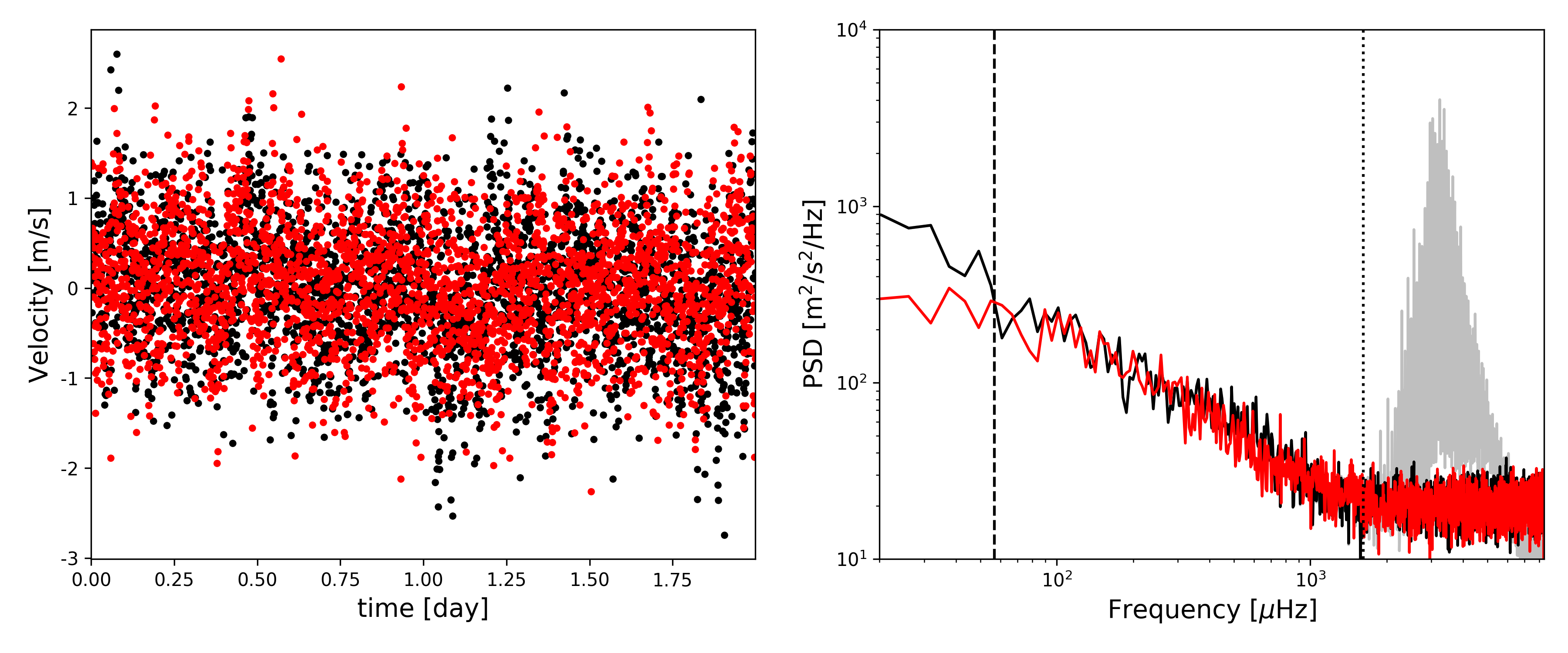}}
        \caption{Left: Comparison of observed GOLF solar velocities (black) and synthetic velocities extracted from 3D simulations of the granulation (red). The acoustics modes have been filtered out using a low-pass filter of $1620$ $\mu$Hz passband and a WGN has been added to both series.
         Right: Associated averaged periodograms computed with $L=26$ time series of  $2$ days duration. The grey PSD shows the averaged periodogram resulting from the unfiltered GOLF observations, where we can see the acoustics modes velocity signatures around $2000-6000$ $\mu$Hz. The dotted line indicates the frequency regime where the high-frequency noise has been artificially added to both time series. The dashed line indicates the frequency limit ($\nu=56$ $\mu$Hz) where the PSD is no longer dominated by the granulation noise.}
        \label{Fig3}
\end{figure*}

\subsubsection{Comparison between RV observations and simulations}
\label{Sec2_compa} 

To compare the synthetic velocity time series extracted from our 3D simulations with the GOLF observations, we added to both datasets a synthetic WGN to replace the high-frequency part of their power spectral density (PSD) that had been filtered out due to the presence of the acoustic modes (see Sec.~\ref{Sec222}). The variance of this high-frequency noise was evaluated using the PSD at $\nu>1620$ $\mu$Hz of the non-filtered GOLF data. This WGN does not affect the lower frequency part of the periodogram.
We note that the influence of the four second exposure time of GOLF has been neglected in the computation of the PSD of the synthetic velocities.

The final comparison of the velocities is shown in the left panel of Fig.~\ref{Fig3} for two selected two-day sequences.
The corresponding (averaged) periodograms (see Eq. \eqref{eq_PL} in Sec.~\ref{Sec32}), resulting from the average of $L=26$ regularly sampled two-day sequences, are shown in the right panel. The third periodogram represents the PSD of GOLF observations before the filtering of the acoustic modes. 
Toward the lower frequencies, we observe the frequency-dependent behavior of the solar granulation in all periodograms. When using high-resolution observations, the RV contribution of the stellar granulation acts as a frequency-dependent noise source that drastically differs from a WGN (characterized by a flat power over all frequencies). 
We observe a good match between the PSDs of the observed and simulated velocities until $\nu=56$ $\mu$Hz (i.e., $\sim 5$ hours) corresponding to the correlation regime dominated by the granulation process (i.e., $\nu \in [50, 1000]$ $\mu$Hz). 
At lower frequencies, the solar PSD becomes dominated by supergranulation and magnetic activity phenomena (spots and plages) and the comparison of these simulations of granulation with observations becomes obsolete (even if the granulation signal affects also the low frequencies of the PSD). We note that supergranules have longer lifetimes and should generate noise correlated over several days. They are not included in our present MHD simulations but they are also reproducible through 3D simulations, although their computing takes a longer time (e.g., \citeads{2009ASPC..415...63S}). 

\section{Detection}
\label{Sec3}

This section presents the considered statistical model and detection tests. Several results detailed in \citetads{2017ITSP...65.2136S} are summarized below for the sake of completeness since Sec. \ref{Sec4} is aimed at validating the theoretical results from this study on real astrophysical data.
The purpose of the approach is to detect (possibly quasi-) periodic components in a stationary colored noise with partially unknown statistics. 
By ``partially'' we mean that a training dataset of this colored noise is available (through MHD simulations). This noise dataset is independent of the observations (see Sec.~\ref{Sec2}). 
In the following, we assume the training data set contains all the noise sources that can affect the dataset under test. We note that currently, the MHD simulations per se cannot reflect the presence of active regions (spots, plages) due to the finite model precision. Hence, this study shows what can be done in the absence of such noise sources or in the situation where activity signatures can be identified by other means and added to the simulation.

\subsection{Hypothesis testing problem}
\label{Sec31}

Let us consider a time series $X(t_j)$ with $N$ points, evenly sampled on times $t_j=j \times dt$ for $j=1,\hdots,N$ with $dt$ the sampling time step. We consider a binary hypothesis problem of the form:
\begin{equation} 
 \left\{ 
 \begin{aligned}
        {\cal H}_0~: X(t_j) &= E(t_j), \\
        {\cal H}_1~: X(t_j) &= R(t_j)+ E(t_j), \\
 \end{aligned}
 \right.
 \label{hyp}
\end{equation}
\noindent where, under the null hypothesis, ${\cal H}_0$, the data contain only the colored noise $E(t_j)$ (of which a training set is available). The noise $E$ is defined as a zero-mean second-order stationary Gaussian noise with unknown power spectral density $S_E$ and absolutely integrable autocorrelation function $r_E$ (see \citetads{2017ITSP...65.2136S}). The alternative hypothesis, ${\cal H}_1$, represents the case where an unknown RV planetary signal $R(t_j)$ is melded with the colored noise. As illustrated, for example in \cite{2016EAS....78..247S}, RV Keplerian signatures can be well approximated by a limited number of pure oscillations :
\begin{equation} 
 R(t_j; \boldsymbol{\theta}_R) = \sum_{q = 1}^{N_s} \alpha_q \sin(2\pi f_q t_j+\varphi_q),  
 \label{eqR}
\end{equation}
where the vector ${\boldsymbol{\theta}_R}$ collects all the unknown amplitudes $\alpha_q\in \mathbb{R}^{+*}$, frequencies $f_q\in \mathbb{R}^{+*}$ , and phases, $\varphi_q \in [0,2\pi[,$ of the $N_s$ sinusoids.
If a star reflects $N_p$ planetary signatures, $N_s$ is, in general, larger than $N_p$. 
The case $N_s\approx N_p$ corresponds to $N_p$ planets with circular orbits and frequencies close to the Fourier grid. In all situations, $N_s$ is much smaller than the number of Fourier frequencies. 

\subsection{Detection approach: a standardized periodogram}
\label{Sec32}


For simplicity, we consider for this section a unit time sampling $dt=1$ and $N$ even.
When the observation sampling is regular, the search of periodic components can be done using the classical periodogram \citepads{1898TeMag...3...13S} defined as:
\begin{equation}
        P(\nu):= \frac{1}{N} ~ \Big| ~ \sum_{j=1}^{N} X(j)~\mathrm{e}^{-{\rm i}2\pi\nu j} \Big|^2.
        \label{eq_P}
\end{equation}
We note that to express $P$ in units of density (m$^2$/s$^2$/Hz), expression \eqref{eq_P} has to be divided by the passband $1/dt$ \citepads[see Eq. (11.6), Chap. 11 of]{Percival_1994}.
This leads us to consider in \eqref{eq_P} a discrete Fourier frequencies defined as:
$$
\nu_k:=\frac{k}{N},~~~~~~\textrm{for}~~~ k=0,\hdots,N-1.
$$
Owing to the hermitian symmetry of the Fourier transform and because we are not interested in the null frequency, below we consider $P(\nu)$ in \eqref{eq_P} only as evaluated on a subset of $\frac{N}{2}-1$ independent Fourier frequencies corresponding to $k\in \Omega:= \{ 1,\hdots, \frac{N}{2}-1\}$.
Asymptotically, $P$ is an unbiased\footnote{$P$ is asymptotically unbiased as ${\mathbb E}~P(\nu_k)= \ S_E(\nu_k) + {\cal{O}}(1/N)$} (but inconsistent\footnote{$P$ is asymptotically inconsistent as $\textrm{Var}~P(\nu_k) = \ S_E(\nu_k)^2 + {\cal{O}}(1/N)$}) estimate of the PSD \citep[see][Theorems~5.2.1~and~5.2.4]{brillinger1981time}. 
The asymptotic distributions of $P$ under both hypotheses are known $\forall k ~ \in ~ \Omega$:
\begin{equation} \centering
 \begin{aligned}
        & P(\nu_k | H_0) \sim \frac{S_E(\nu_k)}{2} \chi^2_2, ~~~ \text{\citep[see][Theorem~5.2.6]{brillinger1981time}}, \\
        & P(\nu_k | H_1) \sim \frac{S_E(\nu_k)}{2} \chi_{2, \lambda_k}^2, ~~~ \text{\citep[see][Corollary ~6.2]{Li_2014}},
 \end{aligned}
 \label{dist_P}
\end{equation}
 with $S_E$ the (unknown) noise PSD and $\lambda_k = \lambda(\nu_k; S_E, {\boldsymbol{\theta}_R})$ a non-centrality parameter. For $N_s$ sinusoidal components involved under ${\cal H}_1$, the expression of this parameter can be found in Eq. (6) of \citetads{2017ITSP...65.2136S}.
We note that if the noise PSD $S_E$ is unknown, the distribution of $P$ given in \eqref{dist_P} is also unknown.

Assuming now that $L$ time series of the colored noise (denoted by $\{X_\ell\}, \ell=1,\hdots, L$ below), can be generated under ${\cal H}_0$ as a training dataset, we propose to use them as an estimate of the noise PSD to calibrate the periodogram of the data under test.
Based on these $L$ time series, we compute an averaged periodogram defined as:
\begin{equation} 
        \overline{P}_L(\nu_k | {\cal H}_0): = \frac{1}{L}~\sum_{\ell=1}^{L}~\frac{1}{N}~\Big| \sum_{j=1}^{N} ~X_\ell(j)~\mathrm{e}^{-{{\rm i}}2\pi\nu_k j} \Big| ^2.
 \label{eq_PL}
\end{equation}
We note that this periodogram has been initially introduced by \citetads{10.2307/2332141}, and used on subseries) to reduce the variance of the classical periodogram given in \eqref{eq_P}.

This averaged periodogram is an asymptotically consistent and unbiased estimator of the PSD. 
Following the same reasoning as for \eqref{dist_P}, the asymptotic distribution of $\overline{P}_L$ can be easily derived $\forall k \in \Omega$ as:
\begin{equation} 
        \overline{P}_L(\nu_k | {\cal H}_0) \sim S_E(\nu_k)~ \frac{\chi_{2L}^2}{2L}.
        \label{dist_PL}
\end{equation}
Using \eqref{eq_PL} to calibrate \eqref{eq_P}, we define the standardized periodogram as:
\begin{equation} 
         \widetilde{P}(\nu_k):= \frac{P(\nu_k)}{\overline{P}_L(\nu_k)}.
         \label{eq_Ptilde} 
\end{equation}

Thanks to the known distributions of the numerator and denominator of \eqref{eq_Ptilde} and to their mutual independence, we can also derive the distribution of the standardized periodogram. Using \eqref{dist_P} and \eqref{dist_PL}, we obtain a ratio of two independent $\chi^2$ variables. This ratio leads, under ${\cal H}_0$ and ${\cal H}_1$, to a central and non-central F-distribution with respectively $2$ and $2L$ degrees of freedom: 
\begin{equation} 
 \begin{aligned}
         & \widetilde{P}(\nu_k| {\cal H}_0) \sim \frac{\chi_2^2/2}{\chi_{2L}^2/2L} \sim F(2,2L), \\
         & \widetilde{P}(\nu_k| {\cal H}_1) \sim \frac{\chi_{2, \lambda_k}^2/2}{\chi_{2L}^2/2L} \sim F_{\lambda_k}(2,2L).
         \end{aligned}
 \label{dist_Ptilde}
\end{equation}
We note that under $\mathcal{H}_0$, the distribution of the standardized periodogram $\widetilde{P}$ is now asymptotically independent of the noise PSD $S_E$. This important property makes tests applied to $\widetilde{P}$ act as \textit{Constant False Alarm Rate} detectors \citep{Scharf94}: their false alarm rate is independent of the noise PSD. This is a very desirable feature in practice since it allows to control the false positive rate despite the unknown noise PSD. Under $\mathcal{H}_1$, the distribution depends on the noise PSD through the non-centrality parameter $\lambda_k$. 
The definition and the analysis of the theoretical performance of tests based on \eqref{eq_Ptilde} are summarized in the following section. 

\subsection{Analysis of tests applied to the standardized periodogram}
\label{Sec33}

Before introducing the tests, it is convenient to consider vectors of random variables, noted in bold. For instance, the vector collecting the periodogram ordinates is written as:
$${\bf{P}}:=[P(\nu_1),\hdots,P(\nu_N)]^\top.$$ 
Notation ${\bf{x}}|{\bf{y}}$ denotes a standardization of the entries of ${\bf{x}}$ by those of ${\bf{y}}$. For instance, the vector of periodogram ordinates is standardized as in \eqref{eq_Ptilde} and defined on the frequency set $\Omega$. \sop{It is} written as:
$$
{\bf{\widetilde{P}\;|\;\overline{P}}}_L:=\left[\frac{P(\nu_1)}{\overline{P}_L(\nu_1)},\hdots, \frac{P(\nu_{\frac{N}{2}-1})}{\overline{P}_L(\nu_{\frac{N}{2}-1})}\right]^\top.
$$

\subsubsection{Test designed for a single periodicity}

 A common test consists of comparing the maximum periodogram value to a detection threshold $\gamma \in \mathbb{R^+}$ that determines the false alarm rate:
\begin{equation}
 T_{M}({\bf{\widetilde{P}\;|\;\overline{P}}}_L):= \displaystyle{\max_k} ~\widetilde{P}(\nu_k) ~ \mathop{\gtrless}_{\mathcal{H}_0}^{\mathcal{H}_1} \gamma.
        \label{tmax}
\end{equation}
This test is most efficient when a single periodicity on the Fourier grid is present under $\mathcal{H}_1$ \citepads{donoho2004}.
As the asymptotic distribution of $\widetilde{P}$ is known at each frequency (see Eq. \eqref{dist_Ptilde}), the false alarm and detection probabilities (noted ${P_{FA}}$ and ${P_{DET}}$ respectively), as well as their relationship ($P_{DET}(P_{FA})$), can be derived analytically \citepads{2017ITSP...65.2136S}:
\begin{equation} 
P_{FA}( \gamma) := \textrm{Pr} \left(T_M({\bf{\widetilde{P}\;|\;\overline{P}}}_L) > \gamma | {\cal H}_0\right) = 1- \Big( 1-\Big(\frac{L}{\gamma+L}\Big)^L\Big)^{N_i},
 \label{pfa}
\end{equation}  
\begin{equation} 
P_{DET}( \gamma) := \textrm{Pr} \left(T_M({\bf{\widetilde{P}\;|\;\overline{P}}}_L) > \gamma | {\cal H}_1 \right)\;\approx\;1\;-\;\displaystyle{ \prod_{k \in \Omega}} \Phi_{F_{\lambda_k}}(\gamma, 2,2L),
 \label{pdet}
\end{equation}  
\begin{equation} 
P_{DET}(P_{FA}) \approx 1 - \displaystyle{ \prod_{k\in \Omega }} \Phi_{F_{\lambda_k}}(\epsilon ,2,2L),
 \label{pfa_and_pdet}
\end{equation}  
where $N_i := \frac{N}{2}-1$ is the number of frequencies effectively considered in the test, $\epsilon := L \Big[ \Big( 1 - \Big( 1 - P_{FA}\Big)^{\frac{1}{N_i}} \Big)^{-\frac{1}{L}} -1 \Big]$ and $\Phi_{F_{\lambda_k}}$ is the cumulative distribution function (CDF) of a non-central F variable with non centrality parameter $\lambda_k$.
The $P_{DET}$ expressions given in \eqref{pdet} and \eqref{pfa_and_pdet} are approximations due to the approximate independence of the periodogram ordinates under ${\mathcal{H}_1}$ \citep[see][Theorem.~6.5]{Li_2014}.
However, the analytic formulae above are quite accurate for values of $N_i$ considered in practice as shown in \citetads{2017ITSP...65.2136S}. These results allow saving a substantial amount of computation time for comparing the tests (in comparison with a MC simulation-based approach). They also allow gaining theoretical insight into the relative performances of the tests.
Using the relation $P_{DET}(P_{FA})$, receiver operating characteristic (ROC) curves can be computed to compare the performances of the statistical tests. Furthermore, these analytical results can also be used to design detectability studies \sop{(see  Sec.~\ref{sec_num2})}.

\subsubsection{Tests designed for multiple periodicities}

Testing for the largest peak in the periodogram may not be the best strategy for the case of multiple (quasi-) periodic signals. 
\citetads{10.2307/2345607} showed that in such cases tests exploiting order statistics of the periodogram may be more powerful than $T_M$ (which looks at the maximum value only). 
In the case where the number of periodogram ordinates at Fourier frequencies affected by the planetary signature can be guessed or estimated a priori (let $N_C$ denote this number), a generalization of test $T_M$ replaces the maximum by the $N_C^\textrm{th}$ largest periodogram components. 
For such a test, analytic expressions for both the $P_{DET}$ and $ P_{FA}$ can also be derived (see test $T_C$ in \citeads{2017ITSP...65.2136S}). 

In practice, however, $N_C^\textrm{th}$ is often unknown and it is necessary to turn to tests that are adaptive with regard to the number of periodicities contained in the total Keplerian signature. Such tests are based on the $P$-values (noted $v$ below) of the standardized periodogram. In the framework considered here, the P-values of an observed random variable (periodogram, or test statistic) is defined as the probability, under the null hypothesis, of obtaining a more extreme value than the observed one. Precisely, the $P$-values of $\widetilde{P}(\nu_k)$ are defined $\forall k ~ \in ~ \Omega$ as:
 $$
        v_{{\bf \widetilde{\bf P}},\;k}:= 1-\Phi_F\left(\widetilde{P}(\nu_k) , 2, 2L\right),
$$
with $\Phi_F$ the CDF of a central F variable.
 Examples of adaptive tests based on the $P$-values are the Higher-Criticism \citepads{donoho2004,2017ITSP...65.2136S} and the Berk-Jones tests 
 \citepads{Berk1979,Aldor2013,Mary2014,Kaplan_2014,Gontscharuk2015,Moscovich_2016} respectively defined as:
\begin{equation} 
        HC({\bf{\widetilde{P}\;|\;\overline{P}}}_L)~~~:=
         \displaystyle{\max_{1 \le { k \le \alpha_0 N}}} 
         \frac{\sqrt{N}(k /N - v_{{\bf{\widetilde{P}},(k)}})}{\sqrt{v_{{\bf{\widetilde{P}},(k)}}(1-v_{{\bf{\widetilde{P}},(k)}})}} ~~~ \mathop{\gtrless}_{\mathcal{H}_0}^{\mathcal{H}_1} \gamma,
 \label{test_HC}
\end{equation}
and
\begin{equation} 
        BJ({\bf{\widetilde{P}\;|\;\overline{P}}}_L)~~~:= \displaystyle{\max_{1 \le k \le \alpha_0 N}}~ I_{1-v_{{\bf{\widetilde{P}},(k)}}} (N-k+1,k) \mathop{\gtrless}_{\mathcal{H}_0}^{\mathcal{H}_1} \gamma,
 \label{test_BJ}
\end{equation}
where $v_{{\bf{\widetilde{P}},(k)}}$ denotes the order statistics of the $P$-values of the standardized periodogram (which have a beta distribution, \citeads{David_2003}), $\alpha_0$ is a constant $\in [\frac{1}{N},1],$ and $I$ denotes the CDF of a beta variable.

Such tests as $HC$ or $BJ$ consist of setting a multiple testing problem, in which a set of test statistics (in our case, this refers to the periodogram at different frequencies) is taken and each of them are simultaneously considered in order to discriminate between the two hypotheses. 
 In essence, these tests compare the maximal deviation of the empirical CDF of the ordered periodogram's $P$-values to their true CDF under $\mathcal{H}_0$. The definition of the deviation depends on the test; both can be seen as variants of a generic divergence \citepads{Zhang2017}. 
 
In periodograms  under ${\mathcal{H}}_1$, the planetary signature affects only a small fraction of the total number of ordinates; furthermore, this is by only a very small amount, leading to a very difficult ``needle in a haystack'' detection problem.
 \citetads{donoho2004} and \citetads{Moscovich_2016} demonstrate theoretically that $HC$ and $BJ$ present optimal guarantees in this regime.
 For finite values of $N_i$, the studies of \citetads{Zhang2017} and \citetads{2017ITSP...65.2136S} show that $BJ$ can be more powerful than other tests in case of weak and non extremely sparse signatures (e.g., multiplanetary systems of small planets with off-grid orbital frequencies and with high eccentricity orbits). \sop{We note that,} in the case of irregular sampling \sop{-- that will be the subject of a second paper}, RV planet signatures can be much less sparse in the Fourier domain than for regular sampling owing to the sidelobes of the spectral window.
Interestingly, efficient and accurate analytic calculations for the distribution of several adaptive tests, such as the $HC$ and $BJ$ under the null and the alternative hypotheses, have been  recently included in \citetads{Zhang2017}. These features make adaptive tests particularly interesting for exoplanets detection, as illustrated in the numerical study below.
 

\section{Numerical study}
\label{Sec4}

In this section, we first evaluate the validity of the statistical method presented in Sec.\ref{Sec3} using the solar observed and synthetic RV time series presented in Sec.~\ref{Sec2}. In a second step, we perform detectability studies for different planet signatures in the presence of solar convective noise by exploiting our analytical results. Finally, we compare the power of classical and adaptive detection tests for different Keplerian signatures.

\subsection{Control of the false alarm: comparison of methods}
\label{sec_num1}

 \begin{figure*}[ht!] \centering
 \resizebox{\hsize}{!}{\includegraphics{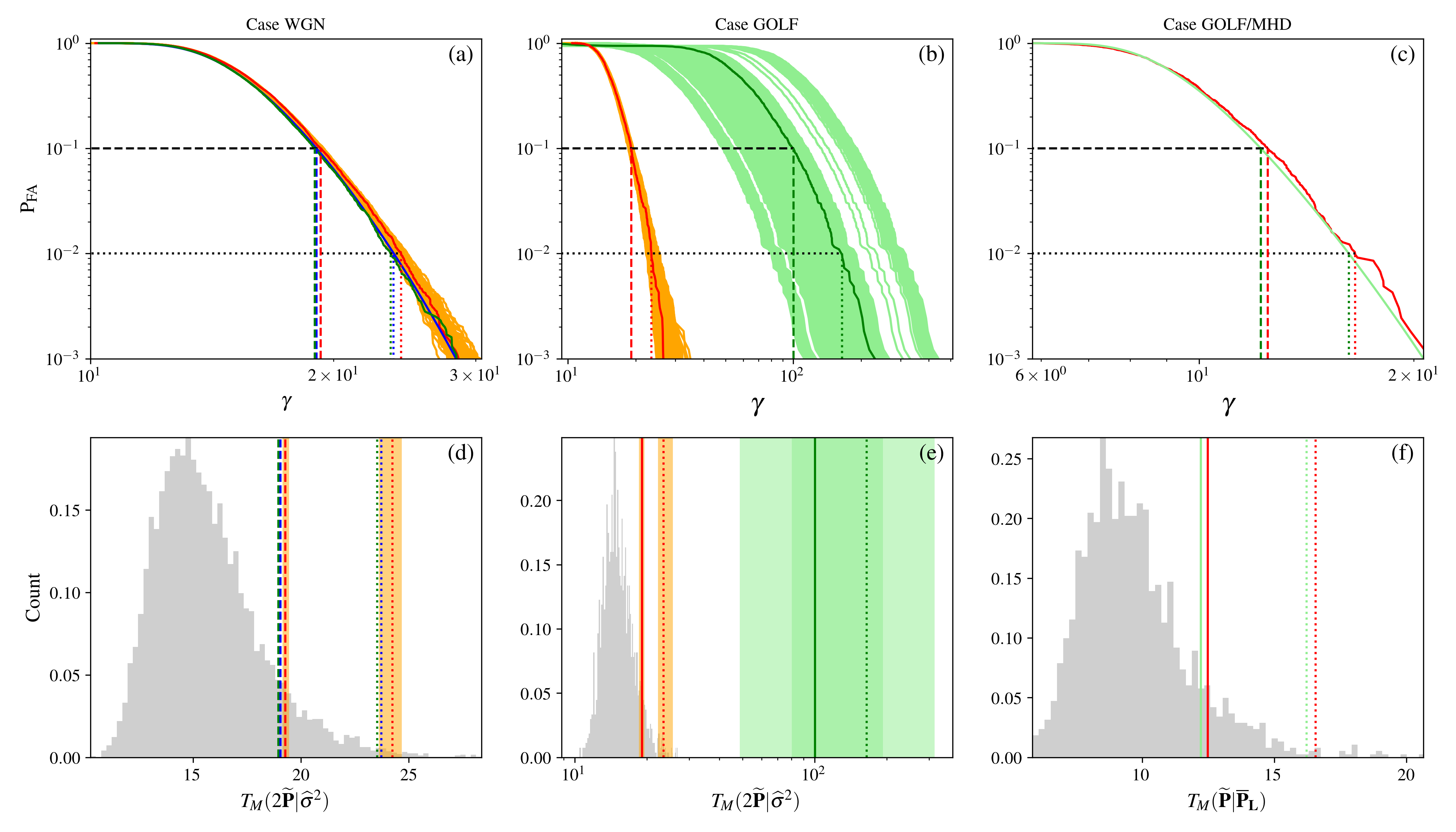}}
        \caption{
Illustration of the reliability of different FAP estimates of test $T_M$ depending on the noise characteristics under ${\cal H}_0$ and on the technique involved.
Top: FAP as a function of the detection threshold $\gamma$ in the case where the data under ${\cal H}_0$ is a WGN of standard deviation $\sigma=49$ cm.s$^{-1}$ (first column) or the colored solar time series of same variance (last two columns).
In panel (a), the blue curve corresponds to the FAP of test $T_M$ with known variance $\sigma^2$ (see Eq. \eqref{eqPhiM}). The red curve corresponds to the FAP of test $T_{M}(2\widetilde{\bf P}|\widehat{\sigma}^2)$ estimated by bootstrap using one estimate of the variance from one time series, $\widehat{\sigma}^2$. The dark green curve represents the true FAP of $T_{M}(2\widetilde{\bf P}|\widehat{\sigma}^2)$. The curves in orange show $100$ FAP estimates of the same test but obtained for $100$ different estimates of $\widehat{\sigma}^2$. 
In panel (b), the red curve shows the FAP of test $T_{M}(2\widetilde{\bf P}|\widehat{\sigma}^2)$, evaluated by bootstrap on one GOLF solar time series with estimated variance $\widehat{\sigma}^2$. The dark green curve shows the true FAP of this test, as estimated using the $N_{series}-1$ other GOLF time series. The orange (resp. light green) curves are the same as the red (resp. the dark green) curves, but using each time a different GOLF time series as input.
In panel (c), the green curve represents the analytic FAP of $T_M$ based on the simulation-standardized periodogram with $L=20$ MHD time series (see Eq.\eqref{pfa}) and the red curve represents the true FAP estimated using $N_{series} = 1640$ GOLF time series. 
Bottom: Empirical distribution of test statistics $T_M$ as estimated by bootstrap (panels (d) and (e)) and by MC simulations of the GOLF series standardized by the MHD simulations (panel (f)). 
In all six panels, the thresholds inferred for FAPs of $1\%$ and $10\%$ by each technique are indicated by the dashed and dotted lines, respectively. The color used for the thresholds in each bottom panel corresponds to the color used for each method in the corresponding upper panel. Numerical values are indicated in Table.~\ref{tab1}.
 }
        \label{fig_appli1a}
\end{figure*} 
 
The first part of this numerical study aims to compare the reliability of different false alarm probability estimates. We compare in particular bootstrap approaches to periodogram standardization (assuming a noise training data set is available). For the sake of concision, we focus on one test: the test of the maximum (see Eq. \eqref{tmax}). As for the considered dataset, we selected the regularly sampled two-day GOLF time series that are available for the first ten years of GOLF observations. In this sample, we removed sequences that are affected by strong outliers due to instrumental defects. 
This corresponds to a set of $N_{series} =1640$ GOLF times series, with $N=2880$ data points each. As described in Sec.~\ref{Sec2}, we filtered out the acoustic modes and added to each time series a WGN of standard deviation $\sigma=49$ cm.s$^{-1}$. 
 This dataset represents our sample of solar observations under $\mathcal{H}_0$, as none of them contains any signs of the Solar System planets (the shortest period, of Mercury, is $\approx 88$ days or $1.31 \times 10^{-7}$ Hz) nor the stellar oscillations modes (affecting mostly the frequencies in the range $1$-$5\times 10^{-3}$ Hz) that have been filtered out. In the following, we will run tests on the frequency range that is dominated by the granulation noise: $\nu \in [50 - 8333] ~\rm\mu$Hz.
 
As discussed in the introduction, a traditional approach in RV planet detection for evaluating FAP thresholds is based on bootstrap procedures.
These methods assume that the observations (or their residuals in the case where some periodicities have been removed) contain only noise and that this noise is further uncorrelated with unknown variance.
The noise statistics are estimated from the observations (see e.g., \citeads{Jenkins2013}, \citeads{Hobson2018}, \citeads{Trifonov2018}, \citeads{Ment2018}).
 The FAP is evaluated by estimating the distribution of the test statistic using fake data, typically obtained by shuffling the data.
 
\begin{table*}[t] \centering
\caption{Threshold values derived in Fig.~\ref{fig_appli1a} for a FAP of $1\%$ and $10\%$ and for our three experiments. The symbol $\dagger$ indicates when thresholds were computed by their sample mean value over a set of MC simulations. 
Note that the disagreement between the values of the last two columns (case GOLF/MHD) is very slight and comes essentially from the limited number of MC simulations used to compute the FAP.}
{\setlength{\extrarowheight}{3pt}
\begin{tabular}{c|c|c|c|c|c|c|c|c|}
\cline{2-8}
& \multicolumn{3}{c|}{ WGN} & \multicolumn{2}{|c|}{GOLF} & \multicolumn{2}{|c|}{GOLF $/$ MHD} \\
\cline{2-8}
& Bootstrap$^\dagger$ & True & $\sigma^2$ known - Eq.~\eqref{pfa_1}& Bootstrap$^\dagger$ & True$^\dagger$ & Eq.~\eqref{pfa} & True \\
\cline{2-8}
\cline{1-8}
\multicolumn{1}{|c|}{$\gamma(P_{FA} = 10\%)$} & 19.04 & 19.1 & 19.05 & 18.9 & 100.3 & 12.21 & 12.50 \\
\multicolumn{1}{|c|}{$\gamma(P_{FA} = 1\%)$} & 23.46 & 24.0 & 23.75 & 23.8 & 162.9 & 16.22 & 16.83\\
\hline
\end{tabular}}
\label{tab1}
\end{table*}

Let us consider first the (scaled) max test, 
\begin{equation}
 T_{M}(2\widetilde{\bf P}|\sigma^2):= \displaystyle{\max_k} ~2~\frac{P(\nu_k)}{\sigma^2} ~ \mathop{\gtrless}_{\mathcal{H}_0}^{\mathcal{H}_1} \gamma,
        \label{tmax2}
\end{equation}
which, by definition, has FAP defined as
\begin{equation}
 P_{FA}(\gamma; T_{M}(2~\widetilde{\bf P}|\sigma^2)\; := \; 1-\Phi_M(\gamma), 
\label{pfa_1}
\end{equation}
where $\Phi_M$ is the CDF of $T_M$.

If the data contains a pure WGN of known variance $\sigma^2$, it can be shown \citepads[see e.g., Sec.~2.4.2,]{sulis:tel-01687077} that
\begin{equation}
\Phi_M= \Big(1-\mathrm{e}^{-\gamma/2}\Big)^{N_i},
\label{eqPhiM}
\end{equation}
with $N_i = N/2-1$ the number of considered (independent) periodogram components.

\sop{In the case where} the variance $\sigma^2$ is unknown, the Max test, taking an estimate of the variance, $\widehat{\sigma}^2$, uses $T_{M}(2\widetilde{\bf P}|\widehat{\sigma}^2)$ as a test statistic.
The bootstrap procedure consists in this case of estimating the variance and repeating the following steps: i) shuffle the observed time series, ii) compute the resulting periodogram on the new data set, and iii) evaluate the test statistics \eqref{tmax2} with $\widehat{\sigma}^2$ replacing ${\sigma}^2$. After generating a large number of realizations of test's statistics, the FAP is derived as in \eqref{pfa_1}, with the empirical distribution $\widehat{\Phi}_M$ replacing ${\Phi}_M$.

This numerical procedure gives good results when the noise is white and Gaussian. 
This is illustrated in panel (a) of Fig.~\ref{fig_appli1a}.
This panel shows three FAP as a function of the detection threshold for the Max test. 
First, the blue line shows the FAP of test $T_{M}(2\widetilde{\bf P}|\sigma^2)$: this is the case for which $\sigma^2$ is known and the FAP is obtained using \eqref{eqPhiM} in \eqref{pfa_1}. 
Second, since the bootstrap procedure describes above estimates $\sigma^2$ for the time series and follows steps i) to iii) above, the estimated function $P_{FA}(\gamma)$ depends on the original data set used to generate the "fake" data set obtained by shuffling. One FAP estimate obtained for one particular data set, a WGN with standard deviation $\sigma = 49$ cm.s$^{-1}$, is shown by the red curve, while the true FAP of $T_{M}(2\widetilde{\bf P}|\widehat{\sigma}^2)$ (evaluated on a WGN of variance $\sigma^2$ instead of $\widehat{\sigma}^2$), is shown by the dark green curve.
Third, if we investigate the dependence of the FAP estimate with the original dataset (used to estimate $\widehat{\sigma}^2$), we obtain the orange curves of the panel (a): here we show $100$ curves corresponding to $100$ different original data sets. 
We see from this panel that the bootstrap procedure is quite stable with \sop{respect} to the considered dataset.

When the noise is colored, the data shuffling breaks the correlations present within the data and the situation changes.
This is shown in panel (b). The red curve shows the FAP of the test $T_{M}(2\widetilde{\bf P}|\widehat{\sigma}^2)$ estimated by bootstrap on one particular time series of estimated variance $\widehat{\sigma}^2$.
The true FAP of this test as estimated from the remaining $N_{series}-1$ is shown by the dark green curve. 
The orange (resp. light green) curves show the same as the red (resp. dark green) curve for all other times series. In contrast to the WGN case, the evaluation of the FAP is not robust nor reliable in this case. Hence, if noise correlations are ignored, a classical bootstrap procedure may severely underestimate the FAP and derive irrelevant thresholds.

This is further illustrated in panels (d) and (e) of Fig.~\ref{fig_appli1a}.
In all bottom panels, the estimated distributions of test $T_M$ are shown in grey. The empirical thresholds corresponding to FA rates of $1\%$ and $10\%$ are represented by the vertical solid and dotted lines, respectively. Their numerical values can be read in Table~\ref{tab1}. 
Panel (d) shows this distribution as obtained from one bootstrap procedure in the case of WGN. In this case, the thresholds estimated by bootstrap are close to the values they should have to ensure the target FAP. However, for the case of solar observations (panel (b)), these estimates are incorrect and lead to FAP that can be an order of magnitude larger than the target value. For instance, for one series (see red curve in panel (b)), the bootstrap procedure derives for a FAP of $1\%$ a threshold value of $\gamma=23.65,$ whereas this value is clearly underestimated: at this threshold, the true FAP, as estimated using all other time series, is in the range of $[70.3\%, 95.1\%]$ (see green light curves).

To conclude this part of the analysis of results on GOLF data, we now turn to test $T_M$ applied to the standardized periodogram (see Eq. \eqref{tmax}). In this case, the theoretical FAP is known (although the noise DSP is analytically unknown) and given by \eqref{pfa}. To verify this expression, we standardize each of the periodograms of the GOLF sequences by the averaged periodogram computed using the $L=20$ noise training datasets generated by the MHD simulations of the granulation (see Sec.~\ref{Sec2}). We then apply test $T_M({\bf{\widetilde{P}\;|\;\overline{P}}}_L)$ and derive the associated $P_{FA}$ as in \eqref{pfa}. The results are shown in panels (c) and (f) of Fig.~\ref{fig_appli1a}. This time, we observe in both cases a very good match between the theoretical FAP and the empirical values (see also last columns of Table.~\ref{tab1}).

We conclude the presentation of this first study with a short discussion. Of course, our point is not to show that the bootstrap is doomed to fail in case of colored noise; rather, it might possible to design bootstrap procedures that would take benefit from a training data set (as the approach of panel (c) does) or would use pre-whitening to obtain more robust FAP estimates than shown in panels (b) and (e). Our point here is primarily to show that noise correlation caused by stellar convection severely impacts FAP estimates and that the proposed approach based on standardization achieves the desired robustness in estimating the FAP.
These results validate the MHD simulation-based standardization approach for the control of the FAP and in particular the accuracy of the analytic calculations for test $T_M({\bf{\widetilde{P}\;|\;\overline{P}}}_L)$ on real data. Since the principle of the approach based on accurate MHD simulations would be unchanged for a different spectral type, these results suggest that it can be used for detecting exoplanets orbiting any type of convective star.

\subsection{Detectability study}
\label{sec_num2}

\begin{figure*}[!ht] \centering
 \resizebox{\hsize}{!}{\includegraphics{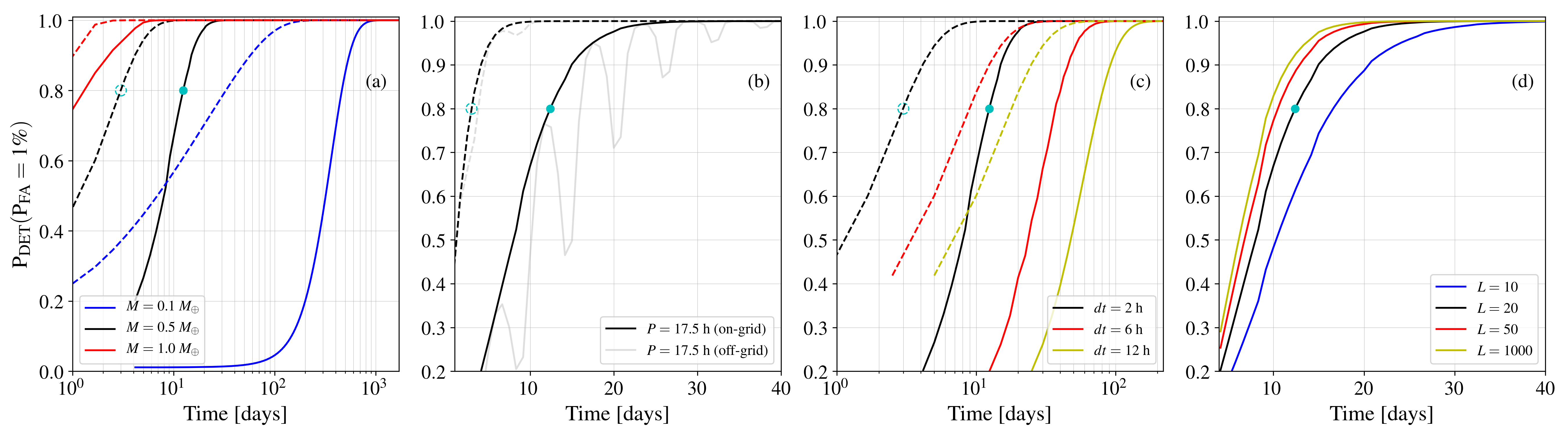}}
\caption{Detection probability as a function of the observation time for a single planet in circular orbit with period $17.5$ hours around a solar-type star, for test $T_M({\bf \widetilde{P}|\overline{P}_L})$, at $P_{FA}=1\%$. The orbital inclination is set to $90$ degrees. 
The different panels show the influence of the planet's mass (a), the orbital period (with the corresponding orbital frequency on (black) or off (gray) -Fourier grid) (b), the time sampling step (c) and the size of the training data set (i.e., the number of available noise times series) (d). In each panel, the black curve indicates the $P_{DET}$ obtained for a configuration in which the planet has a mass of $0.5~M_\oplus$, a circular orbit and $L=20$ HD time series are available for periodogram standardization. The point where $P_{DET}$ reaches $80\%$ for this configuration is indicated by the blue disks. In each panel, the legends indicate the parameters under study. 
 The dashed lines represent the planet's detectability in the case where the noise is white (instead of colored) but with the same standard deviation as the colored convection noise ($\sigma=49$ cm.s$^{-1}$). 
}
\label{fig_detec2}
\end{figure*}

As we have seen in Sec. \ref{Sec3}, exploiting MHD simulations of the granulation noise opens up the possibility for  analytically controlling the false alarm rate and extending the power of the tests for any values of the observation parameters. Comparing the impact of these parameters on the probability of detection for a fixed FAP is very useful in designing observational strategies, for instance. 

Let us consider again the test $T_M$ given in \eqref{tmax}, for which the detection probability can be computed using expression \eqref{pfa_and_pdet} for a given $P_{FA}$. 
Given a specific planetary signature, we want to evaluate the observation duration ($T_{obs}$) that is required to allow for the detection of this planet with a large probability (say, $P_{DET} = 80\%$ at $P_{FA}=1\%$).
We simulated for this study different planetary signatures under ${\cal{H}}_1$ with circular orbits and orbital frequencies on the Fourier grid (we slightly adjusted the time sampling step $dt$ as $T_{obs}$ increases to guarantee that the period is exactly on the grid). For such signatures, only one periodogram ordinate is affected under ${\mathcal{H}_1}$ , while $T_M$ is optimal \citep{donoho2004}. For periodogram standardization, again we used the simulated velocities discussed in Sec.~\ref{Sec2}. The considered convection noise corresponds to a Sun-like star.

Some results are shown in Fig.~\ref{fig_detec2}. 
Each panel of the figure investigates the influence of a different parameter (see legend).
 The black curves correspond to a configuration where a $0.5$ Earth-mass planet orbits circularly its host star with a period of $17.5$ hours, the regular time sampling step $dt$ is $2$ hours and $L=20$ MHD simulations time series are available for periodogram standardization. This setting corresponds to a RV signature of semi-amplitude $K=0.35$ m.s$^{-1}$ and an orbital frequency of $f_p = 1.58 \times 10^{-5}$ Hz. In each panel, the pale blue dot indicates a detection probability of $80\%$ in this configuration for this planet. The dashed lines in the first three panels represent the detectability in the case where the noise is white (instead of colored) but with the same standard deviation as the colored convection noise ($\sigma=49$ cm.s$^{-1}$).
The analytical expression for this probability is (see Eq. (2.53) in \citeads{sulis:tel-01687077}):
\begin{equation}.
 P_{DET}(\gamma; T_{M}(2~\widetilde{\bf P}|\sigma^2)) := 1 - \displaystyle{ \prod_{k \in \Omega}} \Phi_{\chi_2^2,\lambda_k}(\gamma),
\label{pdet_tm2}
\end{equation}
where $\Phi_{\chi_2^2,\lambda_k}$ is the CDF of a non central $\chi_2^2$ distribution with two degrees of freedom and the non-centrality parameter $\lambda_k$ (see Eq. (6) of \citeads{2017ITSP...65.2136S}).

The panel (a) of Fig.~\ref{fig_detec2} shows, for instance, that in the considered configuration, an observation run totaling $T_{obs}\approx 12.4$ days (corresponding to $N\approx150$ sample points with $dt = 2$ hours) would allow the detection of a $0.5$ Earth-mass planet with a probability of $80\%$, while ensuring a false alarm rate of $1\%$. We note that, in contrast, we would only need $T_{obs}=3.0$ days to reach the same trade-off $P_{DET}$ vs $P_{FA}$ if the noise was uncorrelated (see the dashed circle). This factor $\approx 4$ in duration is the price that has to be paid in order to fight against correlation caused by convection noise.
 If the planet mass is lower (panel (a)), the required observational time $T_{obs}$ can increase extensively. For example, for a planet with a mass similar to Mars ($\approx0.1 ~\rm M_\oplus$ leading to an RV semi-amplitude $K=0.07$ m.s$^{-1}$), we would need at least $457$ days of observations to achieve the same performances. Similarly, if the planet's period increases, the needed observational time increases (because the amplitude of the Keplerian signature decreases, which is not shown).
 
 The test's performance depends also on the sampling of the orbital frequency (panel (b)). In our example, if the orbital frequency is not on the Fourier frequency grid, the detection performance of this test decreases, with a loss in $P_{DET}$ that can reach a factor of $2$.

Increasing the sampling time step (panel (c)) or decreasing the number of used training data set (panel (d)), increases also $T_{obs}$. We also note that there is a very small improvement of the test performance brought by increasing $L$ as soon as $L$ is sufficiently large (for $L=50$ and $L=1000$, when the required observation durations are $T_{obs}=10.5$ and $9.5$ days, respectively). This fact is particularly interesting since the MHD simulations are computationally heavy and $L=1000$ may remain outside of the reach of the coming decades.

These plots are examples of false alarm versus power trade-offs that can be achieved by exploiting reliable time series of the convective colored noise. We note that the values indicated in this study are drastically different from those reported in \citetads{2017ITSP...65.2136S}. For instance, we reported $T_{obs}=250$ days for an $1.1~\rm M_\oplus$ planet orbiting its star in $3.2$ days with $dt = 4h$ and $L=100$, while with these same parameters, we find now $T_{obs} \approx 17$ days.
The reason is that the PSD considered to  represent the solar granulation noise source is different from that given in these first works: the considered PSD is now more realistic and deeply checked against Solar observations (see Sec. \ref{Sec2}).

 We now give an example of an application of adaptive tests, which are less well known in the exoplanet community than test $T_M$, although they can sometimes present advantages over the latter. 

\subsection{Detectability of general Keplerian signatures}
\label{sec_num3}

In this section, we compare the performances of the different tests presented in Sec.~\ref{Sec3}, i.e., $T_M$ \eqref{tmax}, HC \eqref{test_HC} and $BJ$ \eqref{test_BJ} for different types of Keplerian signatures. The combination of Keplerian parameters influences the shape of the RV signature which, in turn, influences the sparsity of the signature in the Fourier domain, that is, the number of periodogram components affected by the presence of a planetary signature (e.g., see \citeads{2016EAS....78..247S} for a detailed study of the influence of Keplerian parameters on sparsity).
 Here we define the sparsity coefficient ${\cal S_\beta}$ as the proportion of non-zero coefficients and, as in \citetads{donoho2004}, we parameterize ${\cal S_\beta}$ as:
 $$
{\cal S_\beta} := \frac{N_s}{N} := N^{-\beta},
$$
with $\beta \in [0, 1]$ a sparsity parameter. The value $\beta=1$ corresponds to an extremely sparse signal (i.e., a single periodogram frequency is affected by the periodic signal), and $\beta \to 0$ to a non sparse signature. RV signatures correspond in general to sparse signatures ($\beta$ is typically in the range $[\frac{1}{2}, 1]$). The less sparse signatures are obtained for multiple systems, planets having highly eccentric orbits and planets with off-Fourier grid orbital frequencies. 

\subsubsection{Adaptive tests}

 \begin{figure*} \centering
 \resizebox{0.8\hsize}{!}{\includegraphics{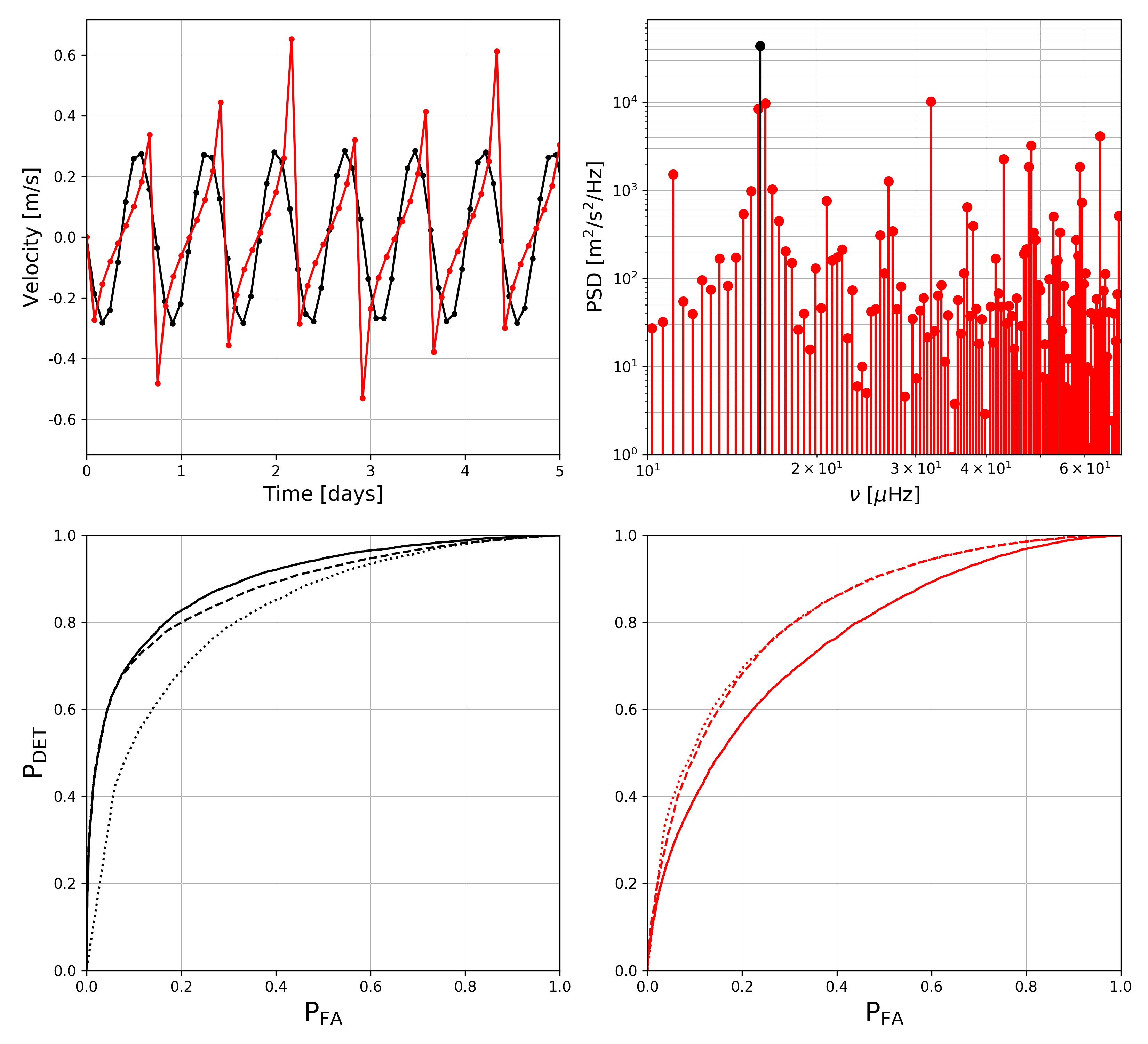}}
\caption{Top: Synthetic RV time series (left) and corresponding periodograms (right) for a single planet in circular (black) or eccentric (red) orbit around a Solar-type star.
For both signals, the planet mass is set to $0.4~M_\oplus$, the period to $17.5$ hours, the orbital inclination to $90$ degrees, the argument at periastron to $\pi/2$ radian, the time sampling step to $2$ hours and the number of data points is $N=300$. For the eccentric planet, the planet orbital frequency is slightly off the Fourier-frequency grid.
Bottom: ROC curves of tests $T_M$ (solid), $HC$ (dashed) and $BJ$ (dotted) applied to ${\bf \widetilde{P}|\overline{P}_L}$ with $L=20$ for the considered circular (left) and eccentric (right) orbital signals.
}
\label{fig_PDET_all_tests}
\end{figure*}

We compare the detection probability of tests $T_M$, $HC$ and $BJ$ for two types of planet signatures.
For the first case, we consider the signal of a $0.4~M_\oplus$ planet in a circular orbit with frequency on the Fourier grid. In the second case, we consider the same planet but with a highly eccentric orbit ($e=0.9$) and a slightly off-grid orbital frequency. The RV signatures of these two planets and the corresponding periodograms are shown in the top panels of Fig.~\ref{fig_PDET_all_tests}. We see a difference in the number of significant peaks in their corresponding periodograms.

We performed MC simulations to evaluate the ROC curves of tests $T_M$, $HC$ and $BJ$. 
As the number of available MHD simulations is limited, we generated the noise under ${\cal H}_0$ using a $20$ order autoregressive (AR) process, with parameters fitted to the MHD simulated time series. 
We generated $10^4$ realizations of this AR process under ${\cal H}_0$ and $10^4$ other realizations under ${\cal H}_1$ with the two RV planetary signatures. For each realization, we computed the standardized periodogram \sop{in Eq. \eqref{eq_Ptilde}} using $L=20$ series \sop{for the denominator} in Eq. \eqref{eq_PL}, and applied the different tests. Results are shown in the bottom panel of Fig.~\ref{fig_PDET_all_tests}.
Comparing the performances of test $T_M$ with the adaptive tests, we observe for the ``sparse signal'' the best results for $T_M$ over the other two tests, with $HC$ close to $T_M$ at low FAP. However, in the case of a high-eccentricity planetary orbit, tests $BJ$ and $HC$ show better performances than $T_M$ at all FAP.
We see that these adaptive tests present another important side
advantage over $T_M$ : their test statistics can be used to reliably estimate the frequency \sop{content} for complex planetary signatures. 

\subsubsection{Ability of adaptive tests in recovering the signal's frequency support}

 \begin{figure*} \centering
 \resizebox{\hsize}{!}{\includegraphics{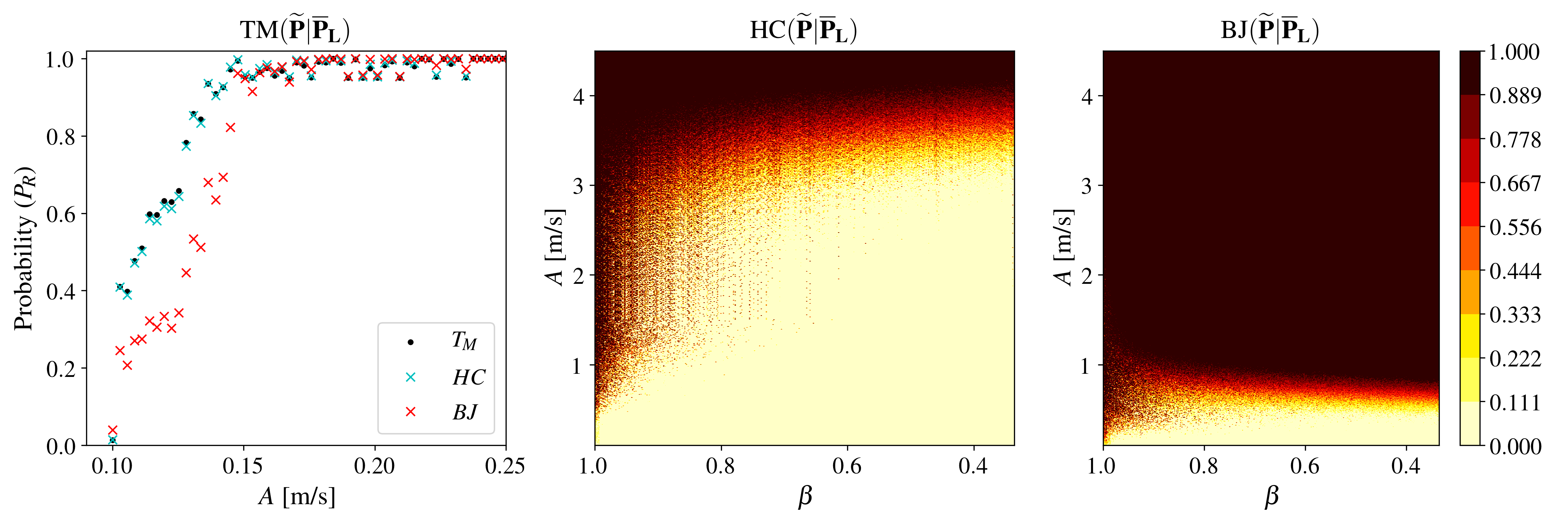}}
\caption{Probability of detecting the correct support (i.e., to find the true number of sinusoids with the correct frequencies) for tests $T_M$ (left), $HC$ (middle) and $BJ$ (right) applied to ${\bf \widetilde{P}|\overline{P}_L}$ with $L=20$. For each given value of $\beta$, the $N_s$ amplitudes are equal. The left panel represents this probability as a function of the sinusoid's amplitude ($N_s=\beta=1$) and the performances of tests $HC$ and $BJ$ have been added for comparison. The two other panels represent the adaptive tests' performances as a function of the sinusoids' amplitude $A$ and the sparsity parameter $\beta$. In these two panels, the probability of correct recovery is indicated in color. }        
        \label{fig_sparsity}
\end{figure*} 
Here we evaluate the ability of each test in detecting the correct number of periodic components at their true location (i.e., the true signal frequencies in the periodogram). We note that this problem is different from that of discriminating between the ``noise only'' vs ``planet plus noise'' hypotheses.
To be able to quantify easily the number of periodic components to be detected under ${\cal H}_1$, we consider  periodic signals in the form of a sum of $N_s$ pure sinusoidal signals with frequencies on the Fourier frequency grid.
Moreover, to generate a large amount of MC simulations, the colored noise is generated as a low order AR process. 
The detection thresholds for a target FAP of $5\%$ were derived for all considered tests by MC simulations on $10^4$ noise sequences under ${\cal H}_0$.
For the training dataset used for periodogram standardization (see Eq. \eqref{eq_Ptilde}), we generated $L=20$ synthetic separate noise time series for each of these $10^4$ sequences. 

In each case, we computed tests $T_M$ \eqref{tmax}, $HC$ \eqref{test_HC}, and $BJ$ \eqref{test_BJ}
on ${\bf \widetilde{P}|\overline{P}_L}$. By definition, test $T_M$ only focuses on the largest component of the (standardized) periodogram. In contrast, tests $HC$ and $BJ$ focus on one particular ordered $P$-value: the one for which the corresponding test statistic is maximum. This ordered $P$-value, say $v_{\widetilde{\bf{P}},i^\star}$ corresponds to the $i^\star$ largest periodogram components. Consequently, when a detection is made, the $i^\star$ largest periodogram ordinates can be used to estimate the signal's frequency support. In the following, we estimate by MC simulations the probability that $i^\star=N_s$ and that the identified frequencies correspond to the true signal frequencies, at
a fixed FAP of $5\%$.

 Under ${\cal H}_1$, we varied the number $N_s$ of sinusoidal signals (by varying parameter $\beta$) added to each of the colored noise time series: $N_s \in [1, 100]$, corresponding to $\beta \in[0.33,1]$. The $N_s$ sinusoids' amplitudes denoted by $A$ in Fig.~\ref{fig_sparsity} were taken as equal, with $A$ varied in the range $[0.1, 4.5]$ m.s$^{-1}$, while their $N_s$ frequency locations $f_{q}$ were picked randomly in the Fourier grid.

The results are shown in Fig.~\ref{fig_sparsity} as a function of the signal parameters (amplitude and sparsity) for tests $T_M$, $HC,$ and $BJ$ (from left to right, respectively).
For $T_M$, we only show the case of $\beta=1$ as this test statistic focuses on the largest periodogram component. 
In the left panel, test $T_M$ appears to be the best to correctly locate the signal frequency when only one frequency is present (the adaptive tests are shown for comparison). For $N_s=1$ ($\beta=1$), the probability of correctly locating the sinusoid frequency ($P_R$) grows faster to $1$ for this test than for the other tests (with $HC$ close to $T_M$).
In contrast to $T_M$, tests $HC$ and $BJ$ allow both for the detection and characterization of the planetary frequency support (middle and right panel of Fig.~\ref{fig_sparsity}). We also note the particularly good performances of $BJ$ vs $HC$ for recovering the frequency support over a large sparsity regime. 
These tests, which benefit from theoretical optimal results (see \citeads{donoho2004} and \citeads{Moscovich_2016}), can be exploitable in the context of exoplanet detection by RV thanks to the considered periodogram standardization. Their good performances in theory and practice make them particularly interesting for the detection and characterization of extrasolar planetary systems.

\section{Discussion}
\label{Sec5}

\subsection{Scope of the proposed method and a zoom on USP planets}

Convective noise affects all components of the periodogram but its effects impact mostly the frequency range corresponding to periods between some minutes and several hours for solar-like stars. 
In Sec.~\ref{Sec2}, we obtained a good match between the observed and simulated PSD of solar RV in this frequency range.
In practice, the convective noise cannot be ``corrected'' as magnetic activity may be (e.g., with chromospheric indicators, \citeads{1995ApJ...438..269B,2018AJ....156..180W}) and constitutes a noise barrier, for which the statistical properties need to be known to reliably claim any planet detection at the cm.s$^{-1}$ level. This was the purpose of our study and its presentation of the formalism of the approach. In a subsequent work, we will apply this formalism to other Solar-like stars having different convective properties.

\begin{figure}[th!] \centering
 \resizebox{\hsize}{!}{\includegraphics{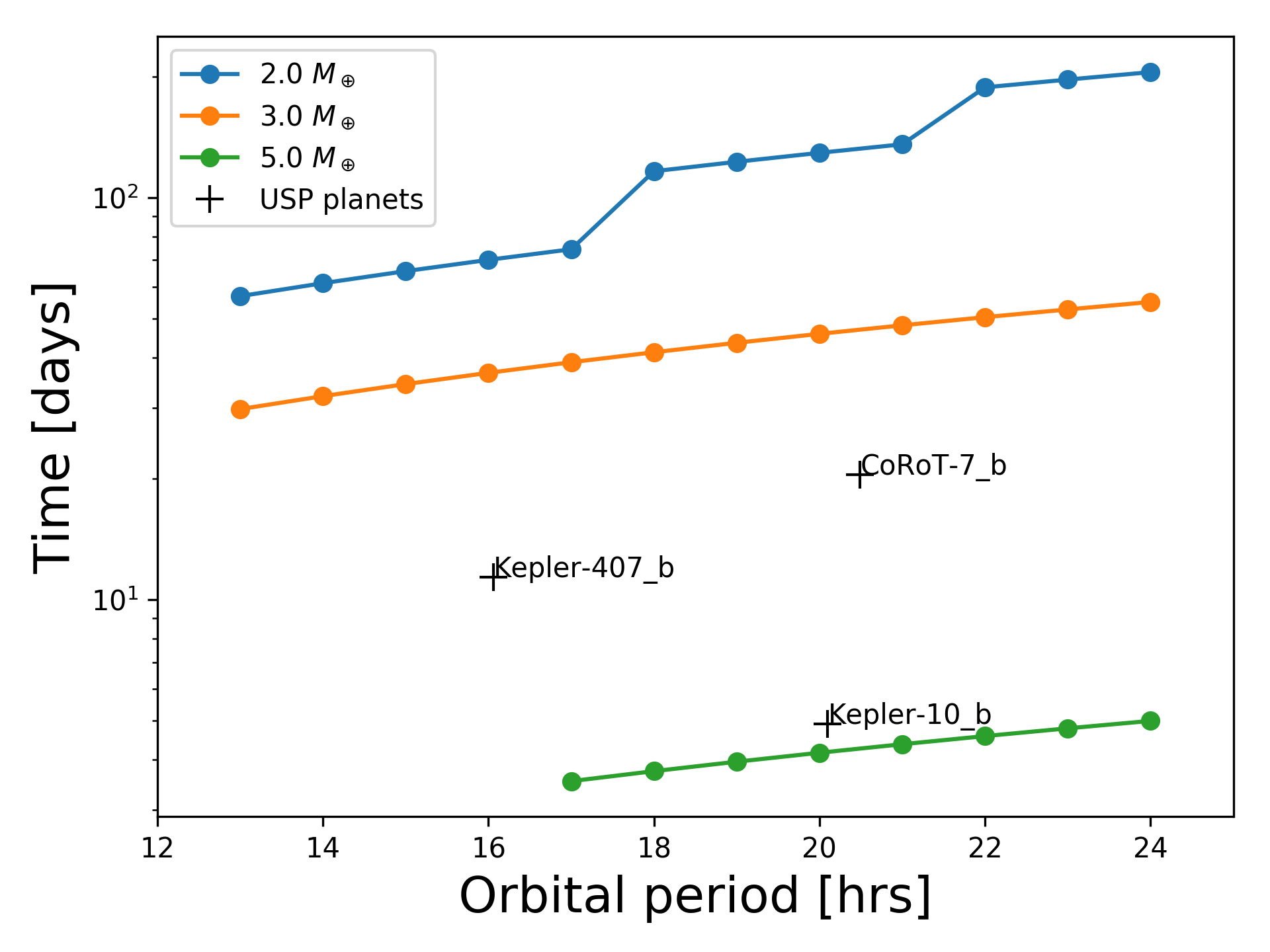}}
\caption{Observation time as a function of the planet orbital period that is needed to achieve $P_{DET}=80\%$ and $P_{FA}=1\%$ with test $T_M$ (colored curves). The fixed parameters are $dt=12$ hrs and $L=2$. The observation time associated with some known USP exoplanets (see Table.~\ref{tab2}, assuming a Solar-like star) are represented by black crosses.}       
        \label{Fig_appli1}
\end{figure}

The proposed method of basing such studies on a standardized periodogram could be directly applied to improve the determination of the FAP in the case of ultra-short period (USP) planets  (defined with periods $<1$ day) under ${\cal H}_1$.
USP planets are known to be tidally locked to their host star (leading to circular orbit) and of small size or mass ($<10M_\oplus$). They exist, in general, in multi-planetary systems.
According to \citetads{Winn_2018}, this category of planets is as frequent as hot-Jupiters (defined with periods ranging up to $10$ days), with one over $200$ Sun-like stars hosting such planets.
%
%
To illustrate the performance of the proposed technique for detecting USP planets, we performed a similar detectability study as described in Sec.~\ref{sec_num2} for some known USP planets (we assume their hosting star is similar to the Sun). 
The results for a sampling rate of $dt=12$ hrs and for $L=2,$ available synthetic noise light curves are reported in Table.~\ref{tab2} and displayed in Fig.~\ref{Fig_appli1}. 
The figure represents the observational time required to reach a $P_{DET}$ of $80\%$ for a  $P_{FA}$ equal to $1\%$, using test $T_M$ given in \eqref{tmax}, as a function of the planet's orbital period. The colored curves show the observational time for virtual planets of different periods and masses. The  black crosses correspond to real planets. Logically, we observe the increase of the observational time with the decrease of the planet mass and the increase of the planet period. 
The table indicates that most USP planets are detectable at the levels specified above with our technique within a couple of weeks ($<21$ days). For small mass USP planets, with $M_p<2M_\oplus$, it would take a couple of months to achieve the same performances. 

Finally,  we  expect the  method presented here to  be  easily  extended to  larger  period  ranges  by computing MHD supergranulation instead of granulation, that is, by extending the simulation domain (making it larger and deeper) with exactly the same simulation setup \citepads{2018LRSP...15....6R}.  Since this would be more demanding in terms of CPU and storage, while retaining, in principle,\ what is shown here, we restrict the scope of this study to granulation scale simulations.

\subsection{Benefits and limitations of the method}

MHD simulations are non parametric, meaning that they do not rely on any adjustable parameter to fit the observed data (see Sec.~\ref{Sec21}). 
In practice, the frequency dependence of the granulation is often estimated using parametric laws, such as Harvey-like profiles \citepads{Harvey_1985}. However, as demonstrated in Sections VI and VII.D of \citetads{2017ITSP...65.2136S}, the estimation of the parameters of these models leads to the injection of an estimation noise in the detection process, the statistics of which are difficult to capture. 
Besides, the  noise parameters derived in this way may be contaminated by the signal to be detected \sop{and the choice of the noise parametric model can be subjective}. 
As we show in this study, using an MHD simulation-based approach allows us to accurately control the estimation noise through the number $L$, whose impact on  the tests performances can be exhibited analytically (see Eqs. \eqref{pdet}-\eqref{pfa}).

This study needs to be extended to other convective stars. The impact of granulation changes throughout the HR diagram: \sop{the larger is the pressure scale height at the surface (that is for larger effective temperatures or lower gravities), the larger are the fluctuations induced by the convective motions.} 
The current limitation in the present method is the computational cost of the MHD simulations to generate a substantially long time series of velocities. However, in the coming years, the increased speed of CPU resources will allow for such computations to be carried out in a more systematic way.  In a subsequent work, we will explore these effects for selected targets in the HR diagram.
Based on the realism of the 3D MHD simulations,  our results suggest that the proposed method can be a powerful and reliable way of detecting RV exoplanet signatures at the cm.s$^{-1}$ level in the presence of convective noise.

A second limitation is the regular sampling involved in this study. 
In practice, the RV data are irregularly sampled and the FAP of any test based on any periodogram \citepads{1898TeMag...3...13S,1982ApJ...263..835S,Zechmeister_2009} cannot be controlled by analytical expressions in the case of correlated noise because the periodogram components are interdependent.\\
However, as mentioned earlier in this paper, we underline that if the irregularity of the considered sampling remains weak, the analytical studies presented here may provide a useful proxy of the tests' performance in practical situations. \sop{For example, this can be used to design detectability studies}. For strongly irregular samplings, the techniques based on the MHD standardized periodogram presented here need to be adapted by dedicated bootstrap procedures (see, i.e., \citetads{2017arXiv170606657S}).
The application of this procedure to the real data deserves a full study that will be the purpose of a second paper.

\begin{table}[t] \centering
\caption{Table of some known USP planets given in the \url{exoplanetarchive.ipac.caltech.edu} catalog. For all targets, the stellar mass is assumed to be $1~M_\odot$ and the eccentricity $0$. Columns indicate the planet's name, orbital period, mass and the observational time we need with the proposed technique to achieve $P_{DET}=80\%$ and $P_{FA}=1\%$ with test $T_{M}({\bf{\widetilde{P}\;|\;\overline{P}}}_L)$ computed for $L=2$ and $dt=12$ hrs. 
Symbols ($\star$) indicate that the target is detectable with probability higher than $80\%$ for the considered parameters (mass and period) and sampling rate. For instance, WASP-47 e would be detectable with probability $>97\%$ as soon as the observation time is superior to $6$ days and 55 Cnc e with a probability $>99\%$.
}
{\setlength{\extrarowheight}{3pt}
\begin{tabular}{|c|c|c|c|}
\hline
Planet name & Period [hrs] & Mass [$M_\oplus$]  & time [days] \\
\hline
CoRoT-7 b & 20.49 & 3.18 & 20.49 \\ 
Kepler-407 b & 16.06 & 3.20 & 11.38 \\
Kepler-10 b & 20.10 & 4.61 & 4.91 \\ 
WASP-47 e & 18.95 & 6.83 & ($\star$) \\
55 Cnc e & 17.68 & 8.08 & ($\star$) \\
\hline
\end{tabular}
}
\label{tab2}
\end{table}

\section{Conclusions}
\label{Sec6}

In cases where the effective temperature, surface gravity, and metallicity of the star are precisely known (thanks to asteroseismology, interferometry, or spectroscopy), 3D MHD simulations are capable of generating realistic RV time series of the stellar granulation. This has been demonstrated for the Sun as part of studies involving the comparison of velocities extracted from 3D spectra of the sodium doublet and GOLF/SoHO observations. 

Following the theoretical analysis described in \citetads{2017ITSP...65.2136S}, we 
used these synthetic time series of the granulation colored noise to design standardized periodograms. These new standardized periodograms allow for the application of tests that are both powerful and for which we can derive accurate FAP.
We present extensive numerical results based on real and synthetic data, including studies on the robustness, the detectability, and the frequency support recovery. In particular, we introduced adaptive tests, which are new in the field of RV planet detection.
Even if the objective of this study is to detect planets down to the cm.s$^{-1}$ level, the proposed procedure is a general approach that can be applied to many periodicity detection problems \sop{in astrophysics} (and beyond).  


\begin{acknowledgements}
This work was supported by the ``Programme National de Physique Stellaire'' (PNPS) of CNRS/INSU co-funded by CEA and CNES. 
S. Sulis acknowledges support from the Austrian Research Promotion Agency (FFG) under project 859724 ``GRAPPA'', as well as Thales Alenia Space and PACA region.
D. Mary acknowledges support from the GDR ISIS through the \textit{Projet exploratoire TASTY}. 
Computations have been done on the ``Mesocentre SIGAMM'' machine, hosted by \textit{Observatoire de la C\^ote d'Azur}.  
 The GOLF instrument onboard SOHO is a cooperative effort of scientists, engineers, and technicians, to whom we are indebted. SOHO is a project of international collaboration between ESA and NASA. 
\end{acknowledgements}

\bibliographystyle{aa} 
\bibliography{Bibfile}

\end{document}